\documentclass[notitlepage,onecolumn,floatfix,preprintnumbers,aps,prd,showpacs,nofootinbib,amsmath,amssymb,amsfonts]{revtex4-1}
\usepackage{bm}
\usepackage[usenames]{color} %\usepackage[dvips,usenames]{color}
\usepackage{dsfont}
\usepackage{graphicx}
\usepackage{hyperref}
\usepackage{textcomp}
\newcommand{\BarChristoffel}[3]{\bar{\Gamma}^{#1}_{\phantom{#1} #2 #3}}

\newcommand{\Lie}{\mathcal{L}}

\newcommand{\be}{\begin{equation}}
\newcommand{\ee}{\end{equation}}
\newcommand{\bea}{\begin{eqnarray}}
\newcommand{\eea}{\end{eqnarray}}

\newcommand{\gr}[1]{{\bm #1}}

\newcommand{\<}{\langle}
\renewcommand{\>}{\rangle}
\newcounter{inout}
\newcommand{\inoutput}{%
	\stepcounter{inout}%
	\theinout}

\newcommand{\xPert}{{\em \lowercase{x}P\lowercase{ert}}}
\newcommand{\xPand}{{\em \lowercase{x}P\lowercase{and}}}

\newcommand{\pert}[2]{{}^{\mbox{\,\tiny $\{#1\}\!$}}{#2}}
\newcommand{\pertbis}[2]{{}^{\mbox{\,\tiny $(#1)\!$}}{#2}}

\newcommand{\mathIn}[1]{
	\begin{quotation} {\em In[\inoutput] :=\ } {\tt #1} \end{quotation}}

\newcommand{\mathInTwoLines}[2]{
	\begin{quotation} {\em In[\inoutput] :=\ } {\tt #1} \\ 
	\phantom{\em In[.] := \quad \;\;} {\tt #2} \end{quotation}}

\newcommand{\mathOut}[1]{
	\begin{quotation} {\em Out[\theinout] :=\ } #1 \end{quotation}}

\newcommand{\mathOutTwoLines}[2]{
	\begin{quotation} {\em Out[\theinout] :=\ } {\tt #1} \\ 
	\phantom{\em Out[\theinout] := \quad \;\,}  {\tt #2}\end{quotation}}
	
\newcommand{\mathOutThreeLines}[3]{
	\begin{quotation} {\em Out[\theinout] :=\ } {\tt #1} \\ 
	\phantom{\em Out[\theinout] := \ \;\,}  {\tt #2}\\
	\phantom{\em Out[\theinout] := \quad \;\,}  {\tt #3}\end{quotation}}

\newcommand{\mathComment}[1]{
	\begin{quotation} \qquad \qquad #1 \end{quotation}}

\newcommand{\ttgr}[1]{{#1}}

\newcommand{\bb}{{\ttgr{\alpha}}}
\newcommand{\cc}{{\ttgr{\beta}}}

\newcommand{\ii}{{\ttgr{\mu}}}
\newcommand{\jj}{{\ttgr{\nu}}}

\begin{document}

\title{\textit{xPand}: An algorithm for perturbing homogeneous cosmologies}

\author{Cyril Pitrou$^{1, 2}$, Xavier Roy$^3$ and Obinna Umeh$^3$}
%\email{cyril.pitrou@port.ac.uk}
\affiliation{
$^1$\mbox{Institut d'Astrophysique de Paris, Universit\'e  Pierre~\&~Marie Curie---Paris VI}, 
\mbox{CNRS-UMR 7095, 98 bis, Bd Arago, 75014 Paris, France} \\ 
$^2$\mbox{Institut Lagrange de Paris, Sorbonne Universit\'es, 98 bis Bd Arago, F-75014 Paris, France}
$^3$\mbox{Astrophysics, Cosmology and Gravity Centre (ACGC), and Department of Mathematics and Applied Mathematics,}
\mbox{Cape Town University, Rondebosch 7701, South Africa}}

%%% Abstract

\begin{abstract}

In this paper, we develop in detail a fully geometrical method for deriving perturbation equations 
about a spatially homogeneous background. This method relies on the $3+1$ splitting of the background space--time 
and does not use any particular set of coordinates: it is implemented in terms of geometrical quantities only, 
using the tensor algebra package \textit{xTensor} in the \textit{xAct} distribution along with the extension for perturbations \textit{xPert}. 
Our algorithm allows one to obtain the perturbation equations for all types of homogeneous cosmologies, 
up to any order and in all possible gauges. As applications, we recover the well-known perturbed Einstein equations 
for Friedmann--Lema\^itre--Robertson--Walker cosmologies up to second order and for Bianchi I cosmologies at first order. 
This work paves the way to the study of these models at higher order and to that of any other perturbed Bianchi cosmologies, 
by circumventing the usually too cumbersome derivation of the perturbed equations.

\end{abstract}

\date{\today}

\maketitle

\vspace{-7mm}

%%% Section
\section*{Introduction}

\vspace{-1mm}

Cosmological perturbation theory constitutes the cornerstone of our current understanding of the origin, 
evolution and formation of large-scale structures. The evolution history of perturbations is written about 
a fixed homogeneous and isotropic Friedmann--Lema\^itre--Robertson--Walker (FLRW) background space--time, 
and the interpretation of cosmological observations (such as WMAP, QUIET and ACT~\cite{Hinshaw:2012fq,Sievers:2013wk,Araujo:2012yh}), 
within this model, converge toward a coherent and unified picture of the Universe. This picture is likely 
to get even clearer when the next generation of large-array cosmological observations (such as EUCLID and SKA) 
becomes operational. It is widely expected that these observations will generate large amount of data that 
will provide a percent level accuracy for the cosmological parameters. 

Cosmological perturbation equations are simple and straightforward to derive at linear order, but they 
are inadequate for understanding the late-time evolution of the Universe, precisely when the 
nonlinear gravitational effects, carrying the information of the physics of current interest, come into play. 
Going beyond first order is a difficult task, and in some cases it becomes extremely arduous to even perform 
a coordinate or gauge transformation at nonlinear order by hand. To the best of our knowledge, there is 
no available easy-to-use software designed for cosmology and capable of deriving all equations of motion 
for perturbed variables. The closely related available option is the GRTensor package \cite{GRTensor}, 
which runs on {\it Maple} or {\it Mathematica}. However, the outputs generated at linear order are 
already a bit complicated to understand, let alone its outputs at nonlinear orders, as it relies exclusively 
on a properly defined set of background coordinates each time it acts on a perturbed variable. 

To fill up this gap, we have developed an algebra package for cosmological perturbation theory, called 
\textit{xPand}~\cite{xPand}, which uses the tools of the tensor
algebra package \textit{xTensor} and an extension for perturbation,
\textit{xPert} \cite{xAct,Brizuela:2008ra}. The {\it xTensor} and
{\xPert} packages are part of the xAct distribution~\cite{xAct} that run on
{\it Mathematica} and
they are available under the General Public License. \textit{xPert} is specifically designed to perform perturbations on arbitrary background 
space--times~\cite{Brizuela:2006ne,Brizuela:2008ra}, but it lacks the
features for specializing to a specific 
background space--time and a specific form for the metric perturbations, as is needed in the case of cosmology. 
In~\cite{Brizuela:2009qd,Brizuela:2010qu} these packages were used to study perturbations about a spherically 
symmetric space--time, more precisely around a Schwarzschild solution of the Einstein field equations.  
The \textit{xPand} package now allows one to derive, in a simple and user-friendly manner, all the necessary 
equations for cosmological perturbation variables, around any homogeneous background space--times, at any 
order and in any gauge. Specifically, the available type of cosmologies cover the Minkowski, FLRW (flat and curved), 
and Bianchi space--times, and the available predefined gauge are: general gauge (no gauge choice), comoving gauge, 
flat gauge, isodensity gauge, Newtonian gauge and synchronous gauge. 

The paper is organized as follows. In section \ref{sec:review_xPert}, we provide a general overview of the mathematical 
framework on which {\xPert } is built. In section \ref{sec:splitting}, we detail the $3+1$ splitting of the background 
manifold into a family of homogeneous hypersurfaces orthogonal to a fundamental observer's velocity. 
In section \ref{sec:pert_fields}, we decompose the perturbed metric with respect to this foliation and define the scalar, vector 
and tensor perturbations. Each of these sections is supplemented by a presentation of the associated implementation in 
\textit{xPand}, by means of several detailed examples. Finally, we summarize and discuss in section \ref{sec:summarize} 
the features and performances of the package.

%%% Section
\section{Perturbations around a general space--time} \label{sec:review_xPert}

In this section we briefly review the algorithm of {\xPert} \cite{Brizuela:2008ra}, which constitutes the basis of our method. 
For more details about perturbation theory in the context of cosmology, we refer the reader to, e.g., \cite{Bruni1997,Nakamura2007}. 

%---Subsection
\subsection{General framework}\label{secGeneral}

Let us consider in what follows a background manifold $\, \overline{\! \cal M}$ along with its perturbed manifold $\cal M$. 
Both are related by means of a diffeomorphism $\phi \colon \, \overline{\! \cal M} \to {\cal M}$. 
Tensorial quantities are thus transported from one manifold to the other with the help of the pull-back $\phi^\star$, 
the push-forward $\phi_\star$, and their respective inverses. 
The metric of the perturbed manifold relates to that of the background as 
\be
	\phi^\star(\gr{g}) 
		= \bar{\bm g} + \Delta[\bar{\bm g}] 
		= \bar{\bm g} + \sum_{n=1}^{\infty} \frac{\Delta^{n}[\bar{\bm g}]}{n!} \, . 
	\label{eq:expansion_g}
\ee
Here and in the sequel, we use boldface symbols for tensorial quantities, an over-bar for background quantities\footnote{
This convention differs from the one adopted in \cite{Brizuela:2008ra}. 
We however opt for this choice as it reflects the standard usage in cosmological perturbation theory.
}, and the notation $\Delta[\bar{\bm T}]$ (resp.\ $\Delta^{n}[\bar{\bm T}]$) for the total (resp.\ $n^{\mathrm{th}}$ order) 
perturbation of a tensor $\gr{T}$. One may prefer to write $\Delta_\phi$ instead of $\Delta$, as the definition of the perturbations 
depends on the diffeomorphism $\phi$, that is on the choice of the gauge. We however choose to omit this reference for the sake 
of clarity, and in order not to burden the notation unnecessarily, we shall moreover use the short-hand: 
$\bm T = \phi^\star(\bm T)$, for any perturbed quantity. 

Unless otherwise specified, when we write down the components of a tensor, these should be understood as expressed 
in a general basis\footnote{
We use Greek letters ($\alpha$, $\beta$, $\mu$, $\nu$, $\rho$, ...)
for space--time indices.
} (this holds equally for the background and perturbed tensors, and for the perturbations). 
Since all perturbation orders live on the background manifold, as they are the result of the pull-back of a tensorial quantity 
living on the perturbed manifold, we shall raise and lower indices using the background metric. We have for instance 
\be
	\pert{n}{h^{\mu \nu}} 
		= \bar g^{\mu \rho} \, \bar g^{\nu \sigma} \pert{n}{h_{\rho \sigma}} \, , 
	\label{eq:updown_pertmet}
\ee
for the $n^{\mathrm{th}}$ order metric perturbations $\! \pert{n}{\bm h} \equiv \Delta^{n}[\bar{\bm g}]$. 

%---Subsection
\subsection{Expansion of the curvature tensors}

%_Subsubsection
\subsubsection{Mathematical framework}

The inverse of the metric tensor is given by the relation 
\be
	{\bm g}^{-1} = \left( \bm{\bar g} +\Delta[\bm{\bar g}] \right)^{-1} \, . 
\ee
Expanding it into 
\be
	{\bm g}^{-1} 
		= \bm{\bar g}^{-1} \sum_{m = 0}^{\infty} (-1)^m \left( {\bm{\bar g}}^{-1} \Delta \left[\bm{\bar g} \right] \right)^m \, , 
\ee
and making use of the definition $\!\! \pert{n}{\bm h} = \Delta^{n}[\bar{\bm g}]$, 
we obtain the $n^{\mathrm{th}}$ order perturbation of ${\bm g}^{-1}$: 
\be
	\Delta^{n}\left[{\left(\bar g^{-1}\right)}^{\mu \nu}\right] 
		= \sum_{(k_i)} (-1)^m \, \frac{n!}{k_1! \dots k_m!} \, 
		\, \pert{k_m}{h^{\mu \zeta_m}} \; \pert{k_{m-1}}{h_{\zeta_m}^{\phantom{\zeta_m} \zeta_{m-1}}} 
		\; \dots \; 
		\pert{k_2}{h_{\zeta_3}^{\phantom{\zeta_3} \zeta_2}} \; \pert{k_1}{h_{\zeta_2}^{\phantom{\zeta_2} \nu}} \, , 
	\label{eq:pert_met}
\ee
where the sum $\sum_{(k_i)}$ runs over the $2^{n - 1}$ sorted partitions of $n$ for $m \le n$ positive integers, 
such that $k_1 + \dots k_m = n$. Note that 
$\Delta^{n} \left[ {(\bar g^{-1})}^{\mu \nu} \right]
\neq \! \pert{n}{h^{\mu \nu} }= \bar g^{\mu \rho} \, \bar g^{\nu \sigma} \pert{n}{h_{\rho \sigma}}$ 
(e.g., we have at first order: $\Delta^1[{(\bar g^{-1})}^{\mu \nu} ] = -\pert{1}{h^{\mu \nu}}$).

By means of relation \eqref{eq:pert_met}, we can express the perturbation of 
the connection components as \cite{Brizuela:2008ra}
\be
	\Delta^{n} \left[ \BarChristoffel{\rho}{\mu}{\nu} \right]
		= \sum_{(k_i)} (-1)^{m+1} \frac{n!}{k_1! \dots k_m!} \, \, 
		\pert{k_m}{h^{\rho \zeta_m}} \; \pert{k_{m-1}}{h_{\zeta_m}^{\phantom{\zeta_m} \zeta_{m-1}}} 
		\; \dots \; 
		\pert{k_2}{h_{\zeta_3}^{\phantom{\zeta_3} \zeta_2}} \; \pert{k_1}{h_{\zeta_2 \mu \nu}} \, , 
	\label{eq:christ_perturb}
\ee
where the last term of the right-hand side is defined by 
\be
	\pert{n}{h_{\rho \mu \nu}} 
		= \frac{1}{2} \left(\bar \nabla_\nu
                  \pert{n}{h_{\rho \mu}} + \bar \nabla_\mu \pert{n}{h_{\rho
                    \nu}} - \bar \nabla_\rho \pert{n}{h_{\mu  \nu}} \right) \, . 
\ee
The perturbation of the Riemann tensor is given in all generality by 
\begin{align}
	\Delta^{n} \left[ \bar R_{\mu \nu \rho}^{\phantom{\mu \nu \rho} \sigma} \right]
		= \bar \nabla_\nu \left( \Delta^{n} \left[ \BarChristoffel{\sigma}{\mu}{\rho} \right]\right) 
		- \sum_{k=1}^{n-1} \binom{n}{k} \, \Delta^k \Big[ 
                \BarChristoffel{\zeta}{\nu}{\rho} \Big] 
		\; \Delta^{n-k} \left[ \BarChristoffel{\sigma}{\zeta}{\mu} \right] 
		\; - \; (\mu \leftrightarrow \nu) \, , 
\end{align}
and for a connection compatible with the metric, we have, from equation \eqref{eq:christ_perturb}, 
\begin{align}
	\Delta^{n} \left[\bar R_{\mu \nu \rho}^{\phantom{\mu \nu \rho} \sigma} \right]
		= & \sum_{(k_i)} \, (-1)^m \frac{n!}{k_1! \dots k_m!} 
	        \bigg( \pert{k_m}{h^{\sigma \zeta_m}} \, \dots \, \pert{k_2}{h_{\zeta_3}^{\phantom{\zeta_3} \zeta_2}}\;
		\bar \nabla_\mu \pert{k_1}{h_{\zeta_2 \rho \nu}} \\
		& + \sum_{s=2}^{m}
		\pert{k_m}{h^{\sigma \zeta_m}} \, \dots \, 
		\pert{k_{s+1}}{h_{\zeta_{s+2}}^{\phantom{\zeta_{s+2}} \zeta_{s+1}}} \; 
		\pert{k_s}{h_{\zeta_s \zeta_{s+1} \mu}}
		\pert{k_{s-1}}{h^{\zeta_s \zeta_{s-1}}} \, \dots \, 
		\pert{k_2}{h_{\zeta_3}^{\phantom{\zeta_3} \zeta_2}}
		\pert{k_1}{h_{\zeta_2 \nu \rho}} \bigg) 
		\; - \; (\mu \leftrightarrow \nu) \, . \nonumber
\end{align}
The symbol $(\mu \leftrightarrow \nu)$ denotes the repetition of the preceding expression with indices $\mu$ and $\nu$ exchanged. 
The perturbation of the Ricci tensor is simply obtained by contracting  the second and fourth indices of 
$\Delta^{n} \left[ \bar R_{\mu \nu \rho}^{\phantom{\mu \nu \rho} \sigma} \right]$ in the previous expression, 
and the perturbation of the Ricci scalar, $\bar R = \bar g^{\rho \sigma} \bar R_{\rho \sigma}$, is written: 
\be
	\Delta^{n} \left[ \bar R \, \right] 
		= \sum_{k=0}^n \binom{n}{k} \, \Delta^{k} \left[ \, \bar g^{\rho \sigma} \, \right] \, 
		\Delta^{n-k} \left[ \bar R_{\rho \sigma} \right] \, . 
\ee
At last, the perturbation of the Einstein tensor is expressed according to 
\be
	\Delta^{n} \left[ \bar G_{\mu \nu} \right] 
		= \Delta^{n} \left[ \bar R_{\mu \nu} \right] 
		- \frac{1}{2} \, \sum_{k=0}^{n} \, \sum_{j=0}^{k} \, \frac{n!}{k! \, j! \, (n-j-k)!} 
		\pert{j}{h_{\mu \nu}} \, \Delta^k \left[ \bar
                  g^{\rho \sigma} \right] \, \Delta^{n-j-k} \left[
                  \bar R_{\rho \sigma} \right] \, . 
\ee

%_Subsubsection
\subsubsection{Implementation in \xPert}

All the perturbative expansions expounded above are implemented in the package \xPert~\cite{Brizuela:2008ra}. 
For completeness, we here briefly review its main commands. 
The package can be loaded by evaluating 
\mathIn{<<xAct`xPert`}
\mathComment{(Version and copyright messages)}
We first define the four-dimensional manifold {\tt M} with abstract indices $\{ \alpha, \, \beta, \, \mu, \, \nu, \, \lambda, \, \sigma \}$: 
\mathIn{DefManifold[ M, 4, $\{ \alpha, \, \beta, \, \mu, \, \nu, \, \lambda, \, \sigma \}$ ];}
and then we define the ambient metric {\tt g} of negative signature, along with its associated covariant derivative {\tt CD}:
\mathIn{DefMetric[ -1, g[-$\alpha$,-$\beta$], CD, \{";","$\bar{\nabla}$"\}, PrintAs->"$\bar {\tt g}$" ];}
where ${\tt M}$ and {\tt g} respectively correspond to $\, \overline{\! \cal M}$ and $\bm{\bar g}$. 
Several tensors related to this metric are automatically defined at the same time (e.g., all the curvature tensors). 
Note that in \textit{xTensor}, the covariant indices of a tensor are represented with a minus sign ({\tt g[-$\alpha$, \!\!-$\beta$]} 
means $\bar g_{\alpha \beta}$), while the latter is omitted for contravariant indices ({\tt g[$\alpha$, \!\!$\beta$]} means $\bar g^{\alpha \beta}$). 

Upon defining the perturbations {\tt dg} of the metric {\tt g} with the command: 
\mathIn{DefMetricPerturbation[ g, dg, $\varepsilon$ ];}
where $\varepsilon$ is the perturbative parameter to be used in the expansions, 
it becomes feasible to evaluate the perturbation of any tensor associated with the metric. 
For instance, the perturbation at first order of the Ricci scalar is simply obtained by evaluating 
\mathIn{ExpandPerturbation@Perturbed[ RicciScalarCD[], 1 ]\,//\,ContractMetric\,//\,ToCanonical}
\mathOut{$R[\bar{\nabla}] - \varepsilon \, dg^{1\bb\cc} R[\bar{\nabla}]_{\bb\cc}
	+\varepsilon\, \bar{\nabla}_\cc \bar{\nabla}_\bb dg^{1\bb\cc} - \varepsilon\,
	\bar{\nabla}_\cc \bar{\nabla}^\cc {dg^{1\bb}}_\bb$}
The functions {\tt ExpandPerturbation} and {\tt Perturbed} are used to evaluate 
the perturbation of any expression up to any order (here at first order), 
the function {\tt ContractMetric} removes the background metric tensor 
through contraction on dummy indices (i.e.\ repeated indices), and the function {\tt ToCanonical} simplifies the result, 
gathering together the terms which are equal up to symmetries. 
Further details can be found in \cite{Brizuela:2008ra}. 

%---Subsection
\subsection{Conformal transformation}

In \textit{xTensor}, the \textit{first} metric defined on the manifold is the one that is used to raise and 
lower any tensor indices. For our purpose, we have chosen it to be the \textit{conformal} metric 
$\bm{\bar g}$ (cf input \textit{In[3]}), which is different from, but conformally related to, 
the background metric of the physical space--time. 
This choice ensures that the conventional way of moving the indices of perturbed fields (equation~\eqref{eq:updown_pertmet}) 
is well satisfied within our algorithm. We however need, now, to relate the tensorial quantities one considers in 
perturbation theory, namely those living on the background manifold of the physical space--time, 
to those we have defined or shall define on the conformal background manifold {\tt M}. 

We detail this important point in the rest of this section.

%_Subsubsection
\subsubsection{Mathematical framework}

Let us denote by $\widetilde{\bm g}$ the metric of the physical space--time, and by 
$\overline{\widetilde{\bm g}}$ its background value. The metrics $\widetilde{\bm g}$ and 
$\bm g$ are related by the conformal transformation 
\be 
	\widetilde g_{\mu \nu} 
		= a^2 g_{\mu\nu} \, , \qquad
	(\widetilde{g}^{\, -1})^{\mu \nu} 
		= a^{-2} (g^{-1})^{\mu \nu} \, , \qquad 
	(\widetilde{g}^{\, -1})^{\mu \rho} \; \widetilde g_{\rho \nu}  
		= \delta^{\mu}_{\phantom{\mu} \nu} \, , 
	\label{eq:def_conf_met}
\ee
with $a$ being the scale factor of the background space--time\footnote{%
We stress again that the first metric defined on {\tt M} is the one that is used to raise and lower any tensor indices. 
Defining another metric, say $\bm f$, on the same manifold from the function {\tt DefMetric} actually creates two objects: 
(i) the tensor $\bm f$, with internal notation {\tt f}, and (ii) the tensor $\bm f^{-1}$, with internal notation {\tt Inv[f]}. 
These have the following properties: 
\be
	f^{\mu \nu} 
		= \bar g^{\mu \rho} \bar g^{\nu \sigma} f_{\rho \sigma} \, , \qquad 
	(f^{-1})^{\mu \nu} 
		= \bar g^{\mu \rho} \bar g^{\nu \sigma} (f^{-1})_{\rho \sigma} \, ,\qquad
	(f^{-1})^{\mu \rho} \, f_{\rho \nu} 
		= \delta^\mu_{\phantom{\mu} \nu} \, , \nonumber
\ee
which explains the notation we use in equation \eqref{eq:def_conf_met}.
}. Substituting into the first expression of \eqref{eq:def_conf_met} the perturbative expansion \eqref{eq:expansion_g} 
for $\bm g$ and its counterpart for $\widetilde{\bm g}$, we extend the conformal transformation to the background: 
\be
	\overline{\widetilde g}_{\mu\nu} 
		= a^2 \, \bar g_{\mu\nu} \, , \qquad 
	(\overline{\widetilde g}{}^{\, -1})^{\mu\nu} 
		= a^{-2} \, \bar g^{\mu\nu} \, , 
	\label{eq:conf_met_back}
\ee
and to the perturbed level: 
\be
	\pert{n}{\widetilde h_{\mu\nu}} 
		= a^2 \pert{n}{h_{\mu\nu}} \, , \quad 
	\quad \mathrm{with} \,\,\,\,
	\pert{n}{\widetilde h^{\mu\nu}} 
		= \bar g^{\mu \rho} \bar g^{\nu \sigma} \pert{n}{\widetilde  h_{\rho \sigma}} \, . 
	\label{eq:conf_met_pert}
\ee
The associated Levi-Civita connections $\bm{\widetilde \nabla}$ and $\bm{\nabla}$, on the one hand, 
and $\bm{\bar{\widetilde \nabla}}$ and $\bm{\bar \nabla}$, on the other, are related by\footnote{%
By definition the scale factor used in the transformation \eqref{eq:def_conf_met} is not to be perturbed. 
Hence we have: $\overline{\widetilde{\bm g}} = \widetilde{\overline{\bm g}}$, and thus: 
$\bm{\bar{\widetilde \nabla}} = \bm{\widetilde{\bar \nabla}}$.
} 
\be
	\widetilde \nabla_\mu \omega_\nu 
		= \nabla_\mu \omega_\nu - C^\rho_{\phantom{\rho}
                  \mu \nu} \omega_\rho \, ,\qquad 
	\bar{\widetilde{\nabla}}_\mu \omega_\nu 
		= \bar{\nabla}_\mu \omega_\nu - \bar C^\rho_{\phantom{\rho} \mu \nu} \omega_\rho \, ,
	\label{eq:rel_cov_derivatives}
\ee
for any 1-form $\bm \omega$. Using equations \eqref{eq:def_conf_met} and \eqref{eq:conf_met_back}, 
we can write the quantities $C^\rho_{\phantom{\rho} \mu \nu}$ and $\bar C^\rho_{\phantom{\rho} \mu \nu}$ as 
\be
	C^\rho_{\phantom{\rho} \mu \nu}
		= 2 \delta^\rho_{\phantom{\rho} (\mu} \nabla_{\nu)} \ln a 
		- g_{\mu\nu} \nabla^\rho \ln a \, ,\qquad 
	\bar C^\rho_{\phantom{\rho} \mu \nu}
		= 2 \delta^\rho_{\phantom{\rho} (\mu} 
		\bar{\nabla}_{\nu)} \ln a - \bar g_{\mu\nu} \bar{\nabla}^\rho \ln a \, , 
	\label{eq:connectors}
\ee
%
% The above set of equations is written in terms of the Nablas. 
%
% Below, they are given in terms of the vector-derivatives, and not in terms of the partial derivatives. 
% Please do not forget that we work in a general basis. 
%
%\be
%	C^\rho_{\phantom{\rho} \mu \nu}
%		= 2 \delta^\rho_{\phantom{\rho} (\mu} {\bf e}_{\nu)} (\ln a) 
%		- g_{\mu\nu} g^{\beta \rho} {\bf e}_\beta (\ln a) \, ,\qquad 
%	\bar C^{\, \rho}_{\phantom{\rho}\mu\nu} 
%		= 2 \delta^\rho_{\phantom{\rho} (\mu} {\bf e}_{\nu)} (\ln a) 
%		- \bar g_{\mu\nu} \bar g^{\beta \rho} {\bf e}_\beta (\ln a) \, ,
%\ee
%
with $\nabla^\rho = g^{\rho \sigma} \nabla_\sigma$ and $\bar{\nabla}^\rho = \bar g^{\rho \sigma} \bar{\nabla}_\sigma$, 
and where the parentheses indicate symmetrization over the indices enclosed. 
We can then formulate the $n^{\mathrm{th}}$ order perturbation of ${C^\rho}_{\mu\nu}$ as\footnote{% 
%
% Beginning of footnote
%
The background and perturbed connections, on the one hand, and the background and conformal connections, on the other hand, 
are respectively related by 
\be
	{\nabla}_\mu {\omega}_\nu 
		= \bar{\nabla}_\mu {\omega_\nu} - \Delta[\bar\Gamma^{\rho}_{\phantom{\rho} \mu \nu}]{ \omega}_\rho \,\, , \qquad 
	\widetilde{\bar \nabla}_\mu \widetilde{\omega}_\nu 
		= \bar{\nabla}_\mu {\widetilde{\omega}_\nu} - \bar C^{\rho}_{\phantom{\rho} \mu \nu} \widetilde{\omega}_\rho \, , \nonumber 
\ee
for any 1-form $\bm \omega$. Performing a conformal transformation on the former expression and perturbing the latter respectively yields 
\bea
	\widetilde{\nabla}_\mu \widetilde{\omega}_\nu 
		& = & \widetilde{\bar{\nabla}}_\mu \widetilde{\omega}_\nu 
			- \widetilde{\Delta[\bar\Gamma^{\rho}_{\phantom{\rho} \mu \nu}]} \, \widetilde{\omega}_\rho 
		= \bar{\nabla}_\mu \widetilde{\omega}_\nu - \bar C^{\rho}_{\phantom{\rho} \mu \nu} \widetilde{\omega}_\rho 
			- \widetilde{\Delta[\bar\Gamma^{\rho}_{\phantom{\rho} \mu \nu}]} \, \widetilde{\omega}_\rho \, , \nonumber \\
	\widetilde{\nabla}_\mu \widetilde{\omega}_\nu 
		& = & {\nabla}_\mu \widetilde{\omega}_\nu - C^{\rho}_{\phantom{\rho} \mu \nu} \widetilde{\omega}_\rho 
		= \bar{\nabla}_\mu \widetilde{\omega}_\nu - \Delta[\bar\Gamma^{\rho}_{\phantom{\rho} \mu \nu}] \, \widetilde{ \omega}_\rho 
			- C^{\rho}_{\phantom{\rho} \mu \nu} \widetilde{\omega}_\rho \, . \nonumber 
\eea
These relations can only be compatible if
%From relations \eqref{eq:rel_cov_derivatives} we can write: 
\be
	\widetilde{\Delta[\bar{\Gamma}^\rho_{\phantom{\rho} \mu\nu}]} -\Delta[\bar {\Gamma}^\rho_{\phantom{\rho} \mu\nu}] 
		= \Delta[\bar C^\rho_{\phantom{\rho} \mu\nu}] \, , \nonumber 
\ee
It can be checked directly from
equations~(\ref{eq:christ_perturb}), (\ref{eq:conf_met_pert}) and (\ref{eq:pert_connector}) that this
is indeed the case.
%and thus we recover the identity of the two previous expressions. 
This shows the equivalence between the transformations $\bar{\bm \nabla} \to {\bm \nabla} \to \widetilde{\bm \nabla}$ and 
$\bar{\bm \nabla} \to \widetilde{\bar{\bm \nabla}} = \bar{\widetilde{\bm \nabla}} \to \widetilde{\bm \nabla}$. 
The latter approach actually proves itself to be faster within \textit{xPand}. 
It is therefore the one that we have coded in the function {\tt ToxPand} (see section \ref{subsec:min_ex}). 
%
% End of footnote
%
}
\be
	\Delta^{n} \left[ \bar C^\rho_{\phantom{\rho}\mu\nu} \right] 
		= \sum_{k=0}^n \, \frac{n!}{k! \, (n-k)!} \pert{k}{h_{\mu\nu}} \, 
		\pert{n-k}{h^{\rho \sigma}} \, \bar \nabla_\sigma \ln a \, . 
	\label{eq:pert_connector}
\ee
With the help of the two previous relations, we are now able 
to provide the correspondence we seek. 

For instance, the Riemann tensor associated with the metric $\overline{\widetilde{\bm g}}$ is given by 
(see appendix D of \cite{1984ucp..book.....W}) 
\begin{align}
	\bar{\! \widetilde{R}} {}_{\mu \nu \rho}^{\phantom{\mu \nu \rho} \sigma} 
		& = \bar{R}_{\mu \nu \rho}^{\phantom{\mu \nu \rho} \sigma} 
		- 2 \, \bar\nabla_{[ \mu} \bar C^\sigma_{\phantom{\sigma} \nu ] \rho} 
		+ 2 \, \bar C^{\zeta}_{\phantom{\zeta} \rho [ \mu} \, \bar C^{\sigma}_{\phantom{\sigma} \nu ] \zeta} \nonumber \\ 
		& = \bar R_{\mu \nu \rho}^{\phantom{\mu \nu \rho} \sigma} 
		+ 2 \, \delta^{\sigma}_{\phantom{\sigma} [ \mu} \bar\nabla_{\nu ]} \bar\nabla_\rho \ln a 
		- 2 \, \bar g_{\rho [ \mu} \bar\nabla_{\nu ]} \bar\nabla^\sigma \ln a \nonumber \\ 
		& \hspace{1.5cm} 
		- 2 \, \delta^{\sigma}_{\phantom{\sigma} [ \mu} \bar\nabla_{\nu ]} \ln a \, \bar\nabla_\rho \ln a 
		+ 2 \, \bar g_{\rho [ \mu} \bar\nabla_{\nu ]} \ln a \, \bar\nabla^\sigma \ln a 
		- 2 \, \bar g_{\rho [ \mu} \delta^{\sigma}_{\phantom{\sigma} \nu ]} \bar\nabla^\zeta \ln a \, \bar\nabla_\zeta \ln a \, , 
	\label{eq:riemann_back_phys}
\end{align}
where the brackets indicate anti-symmetrization over the indices enclosed. 
Perturbing this expression and using equation \eqref{eq:pert_connector}, we can finally relate 
$\Delta^{n}\big[ \, \bar{\! \widetilde{R}} {}_{\mu \nu \rho}^{\phantom{\mu \nu \rho}\sigma} \big]$ to $\Delta^{n}\left[\bar{R}{}_{\mu \nu \rho}^{\phantom{\mu \nu \rho} \sigma}\right]$ and recover the usual quantities studied in perturbation theory. 

%_Subsubsection
\subsubsection{Implementation in \xPand}

The \textit{xTensor} package provides the tools to define a metric conformally related to another, 
thanks to the option {\tt ConformalTo} of the function {\tt DefMetric}. 
We have encapsulated this in \textit{xPand} within the function {\tt Def\-Conformal\-Metric}, 
which furthermore ensures the transitivity of several conformal transformations. 

Let us load the package \textit{xPand}: 
\mathIn{<<xAct`xPand`}
\vspace{-2mm}
\mathComment{- - - - - - - - - - - - - - - - - - - - - - - - - - - - - - - - - - - - - - - - - - -}
\vspace{-4mm}
\mathComment{Package xAct`xPand`  version 0.4.0, \{2013,02,08\}}
\mathComment{CopyRight (C) 2012-2013, Cyril Pitrou, Xavier Roy and Obinna Umeh under the GPL.}
%\mathComment{- - - - - - - - - - - - - - - - - - - - - - - - - - - - - - - - - - - - - - - - - - -}
%
By evaluating the command 
\mathIn{DefConformalMetric[ g, a ];}
we define the scalar factor {\tt a[]} and the metric {\tt ga2} conformally related to {\tt g} through {\tt a} 
({\tt ga2} thus corresponds to the background metric $\overline{\widetilde{\bm g}}$ of the physical space--time). 
To obtain the expression of any tensorial quantities living on the manifold described by {\tt ga2} in terms 
of those defined on {\tt M}, one then simply has to use the \textit{xPand} function {\tt Conformal}. 

For instance, we obtain for the Riemann tensor associated with {\tt ga2} the expression 
\mathIn{Conformal[ g, ga2 ][ RiemannCD[-$\alpha$,-$\beta$,-$\mu$,$\, \nu$] ]}
\mathOutTwoLines{$ \displaystyle 
	R \left[ \bar \nabla\right]{}_{\bb\cc\ii}^{\phantom{\bb\cc\ii}\jj}
- \delta_\cc^{\phantom{\cc} \jj} \, \bar g_{\bb\ii} \, \frac{\bar \nabla_{\lambda} \, a \, \bar \nabla^\lambda \, a}{a^2} 
+ \delta_\bb^{\phantom{\cc} \jj} \, \bar g_{\cc\ii} \, \frac{\bar \nabla_\lambda \, a \, \bar \nabla^\lambda \, a}{a^2} 
+ 2 \, \delta_\cc^{\phantom{\cc} \jj} \frac{\bar \nabla_\bb \, a \, \bar \nabla_\ii \, a}{a^2} 
- 2 \, \delta_\bb^{\phantom{\cc} \jj} \frac{\bar \nabla_\cc \, a \, \bar \nabla_\ii \, a}{a^2} $} 
{$ \displaystyle
- \delta_\cc^{\phantom{\cc} \jj} \frac{\bar \nabla_\ii \bar \nabla_\bb \, a}{a} 
+ \delta_\bb^{\phantom{\cc} \jj} \frac{\bar \nabla_\ii \bar \nabla_\cc \, a}{a} 
- 2 \, \bar g_{\cc\ii} \, \frac{\bar \nabla_\bb \, a \, \bar \nabla^\jj a}{a^2}
+ 2 \, \bar g_{\bb\ii} \, \frac{\bar \nabla_\cc \, a \, \bar \nabla^\jj \, a}{a^2} 
+ \bar g_{\cc\ii} \, \frac{\bar \nabla^\jj \bar \nabla_\bb \, a}{a} 
- \bar g_{\bb\ii} \, \frac{\bar \nabla^\jj \bar \nabla_\cc \, a}{a} $}
which coincides with equation \eqref{eq:riemann_back_phys}. 

The conformal transformation may as well be performed on quantities that are not related to a metric. 
For a general tensor, it is defined as 
\be
	\widetilde{T}^{\mu_1\dots \mu_p}_{\phantom{\mu_1\dots \mu_p} \nu_1\dots \nu_q} 
		= a^{q - p + W(\bm T)} \, T^{\mu_1\dots \mu_p}_{\phantom{\mu_1\dots \mu_p} \nu_1\dots \nu_q} \, , 
\ee
where $W(\bm T)$ is the \textit{conformal weight} of the tensor $\bm T$. The default value of $W$ is chosen to be zero 
in order to leave the norm invariant under a conformal transformation: 
\mathIn{DefTensor[ W[-$\alpha$], M ];}
\mathIn{Conformal[ g, ga2 ][ W[-$\alpha$] ]}
\mathOut{$ a \, W_\bb $}
\mathIn{Conformal[ g, ga2 ][ W[$\alpha$] ]}
\mathOut{$\displaystyle \frac{W^\bb}{a} $}
The conformal weight can however be modified for each tensor with the \textit{xPand} function {\tt ConformalWeight}. 
This can be of use for instance to preserve the geodesic character of light-like vectors\footnote{%
We will make use of this prescription in a future version of \textit{xPand} to implement the derivation of the (perturbed) null geodesic equation.
}:
\mathIn{DefTensor[ k[-$\bb$], M ];}
\mathIn{ConformalWeight[ k ] \hspace{-3mm} \^{} \hspace{-3mm} = -1;}
\mathIn{ConformalWeight[ k[-$\bb$] ]}
\mathOut{$ 0 $}
\mathIn{ConformalWeight[ k[$\bb$] ]}
\mathOut{$ -2 $}
\mathIn{Conformal[ g, ga2 ][ k[$\bb$] CD[-$\bb$]@k[-$\cc$] ]}
\mathOut{$ \displaystyle \frac{k^\bb \, \bar \nabla_\bb \, k_\cc}{a^2} - k_\bb \, k^\bb \, \frac{\bar \nabla_\cc \, a}{a^3} $}

So far, by applying {\tt Conformal} then {\tt ExpandPerturbation@Perturbed} on a given expression defined on {\tt M}, 
one obtains the perturbation of its conformal transformation in terms of the tensors defined on {\tt M}, the metric {\tt g}, 
its perturbations {\tt dg}, the connection $\tt \bar\nabla$, and the scale factor {\tt a}. 
To end up with the usual expressions of perturbation theory, one needs to perform a $3+1$ splitting of the background 
manifold (section~\ref{sec:splitting}), then decompose each perturbed fields into its spatial and temporal parts and 
finally parameterize the perturbations of the metric (section~\ref{sec:pert_fields}).

%%% Section
\section{$3+1$ splitting of the background manifold} \label{sec:splitting}

%---Subsection
\subsection{Induced metric}

The assumption that the background space--time possesses a set of (three-dimensional) homogeneous surfaces 
provides a natural choice for the $3+1$ slicing. We foliate the background manifold by means of this family, 
and we denote by $\bm{\bar n}$ the unit time-like vector ($\bar n^\mu \bar n_\mu = -1$) normal to it. 
The metric of $\, \overline{\! \cal M}$ is decomposed as 
\be \label{eq:def_h}
	\bar g_{\mu\nu} 
		= \bar h_{\mu\nu} - \bar n_\mu \bar n_\nu \, , 
	\qquad \mathrm{with} \qquad 
		\bar h_{\mu \nu} \bar n^\mu = 0 
	\qquad \mathrm{and} \qquad 
		\bar h^\mu_{\phantom{\mu} \rho} \bar h^\rho_{\phantom{\rho} \nu} = \bar h^\mu_{\phantom{\mu} \nu} \, ,
\ee
where $\bm{\bar h}$ represents the induced metric of the spatial hypersufaces\footnote{%
For the sake of clarity, let us note that $\! \pert{n}{\bm h}$ are \textit{not} the perturbations of $\bm{\bar h}$. 
From the definition of $\! \pert{n}{\bm h}$ together with relation \eqref{eq:def_h}, we can actually relate them as 
$\Delta^n[\bm{\bar h}] = \! {}^{\{n\}} \bm{h} - \Delta^n[\bm{\bar n} \otimes \bm{\bar n}]$. 
%
% We have chosen to use these symbols in order to follow the usual notation of both perturbation theory and the $3+1$ formalism. 
%
}. 

The acceleration of the so-called Eulerian observers satisfies in all generality~\cite{Gourgoulhon:2012ue} 
\be
	\bar a_\mu = \bar n^\rho \, \bar \nabla_\rho \bar n_\mu 
		= \frac{\bar D_\mu \bar \alpha}{\bar \alpha} \, , 
	\label{eq:acceleration}
\ee
with $\bar \alpha$ being the lapse function. $\bar{\bm D}$ stands for the connection of the three-surfaces associated with $\bm{\bar h}$ 
($\bar D_\rho \bar h_{\mu \nu} = 0$), and it is related to the four-covariant derivative as 
\be
	\bar D_\rho  T_{\mu_1 \dots \mu_p} 
		= \bar h^\sigma_{\phantom{\sigma} \rho} \bar h^{\nu_1}_{\phantom{\nu_1} \mu_1} \dots \bar h^{\nu_p}_{\phantom{\nu_p} \mu_p} 
		\bar \nabla_\sigma T_{\nu_1 \dots \nu_p} \, , 
	\label{eq:cdtoCD}
\ee
for any spatial tensor field\footnote{%
We recall that the operator $\bar{\bm D}$ loses its character of derivative when it is applied to non-spatial tensors. 
More precisely, we are not allowed to use the Leibniz rule anymore, as one can realize upon writing for instance 
\begin{align*}
	\bar D_\rho ( \psi \, \bar T_{\mu_1 \dots \mu_p} ) 
		& = \psi \bar D_\rho \bar T_{\mu_1 \dots \mu_p} 
		+ \bar h^{\nu_1}_{\phantom{\nu_1} \mu_1} \dots \bar h^{\nu_p}_{\phantom{\nu_p} \mu_p} 
		\bar T_{\nu_1 \dots \nu_p} \bar D_\rho \psi \\ 
		& \neq \psi \bar D_\rho \bar T_{\mu_1 \dots \mu_p} + \bar T_{\mu_1 \dots \mu_p} \, \bar D_\rho \psi \, , 
\end{align*}
for any scalar field $\psi$. 
}. Since the lapse is homogeneous in the configuration at stake, the acceleration vanishes and the observers are in geodesic motion. 
We can therefore label each hypersurface by their proper time $\eta$ and write: $\bar n_\mu = - \bar \nabla_\mu \eta$. 
In addition, $\bm{\bar n}$ being hypersurface-forming by construction, its vorticity vanishes; this property yields 
\be \label{eq:vorticity}
	\bar \omega_{\mu \nu} 
		= \bar h^\rho_{\phantom{\rho} \mu} \bar h^\sigma_{\phantom{\sigma} \nu} \bar \nabla_{[ \rho} \bar n_{\sigma ]} 
		= 0 
	\quad \Leftrightarrow \quad 
	\bar \nabla_{[\mu} \bar n_{\nu]} = 0 \, , 
\ee
where the equivalence stems from the null acceleration. 
For comprehensive reviews on the $3+1$ formalism, we refer the reader to, e.g., \cite{Smarr, Gourgoulhon:2012ue}.
%
% This background $3+1$ splitting can be understood as a particular case of
% the $1+3$ formalism (see Refs.~\cite{Ellis:1971pg,Ellis:1998ct}), or a
% particular case of the general $3+1$ formalism (see Ref.~\cite{Gourgoulhon:2012ue} for a review). 
%

%---Subsection
\subsection{Extrinsic curvature} \label{subsec:extr_curv}

Another tensor we shall make use of is the symmetric extrinsic curvature tensor, 
which characterizes the way the three-surfaces are embedded into the background manifold. 
It satisfies the relation 
\be
	\bar K_{\mu\nu} 
		= \bar h^{\rho}_{\phantom{\rho} \mu} \bar h^{\sigma}_{\phantom{\sigma} \nu} \bar \nabla_\rho \bar n_\sigma \, , 
	\label{eq:ext_curv}
\ee
where we have chosen a positive sign for the right-hand side\footnote{%
This convention does not affect the $3+1$ Einstein equations as written 
in terms of the kinematical quantities of the Eulerian observers.
}. 
From the decomposition \eqref{eq:def_h} along with the vanishing of the 
acceleration $\bm{\bar a}$ and the unitary of $\bm{\bar n}$, we can reformulate expression \eqref{eq:ext_curv} as 
\be
	\bar K_{\mu\nu} = \bar \nabla_{\mu} \bar n_{\nu} \, . 
\ee
Since the volume expansion of the background space--time is entirely contained in the scale factor $a$, 
and owing to the conformal transformation \eqref{eq:conf_met_back}, the trace of the extrinsic curvature vanishes: 
$\bar K^{\mu}_{\phantom{\mu}\mu} = 0$. As a result, we have for general Bianchi cosmologies: 
$\bar K_{\mu\nu} = \bar \sigma_{\mu\nu}$, with $\bar \sigma_{\mu\nu}$ being the shear of the Eulerian observers; 
and for FLRW cosmologies: $\bar K_{\mu\nu} = 0$. 

%---Subsection
\subsection{Curvature tensors}

The splitting of the four-Riemann tensor can be constructed from its different projections onto the spatial slices 
and the congruence of the observers. It is written as 
\begin{align}
	\bar R_{\mu \nu \rho \sigma} = 
		& \, {}^3 \! \bar{R}_{\mu \nu \rho \sigma} + 2 \bar K_{\mu [\rho} \bar K_{\sigma ] \nu}
		- 4 \left( \bar D_{[ \mu} \bar K_{\nu ] [ \rho} \right) \bar n_{\sigma ]}
		- 4 \left( \bar D_{[ \rho} \bar K_{\sigma ] [ \mu} \right) \bar n_{\nu ]} 
		+ 4 \, \bar n_{[ \mu} \, \bar K_{\nu ]}^{\phantom{\nu} \zeta} \, \bar K_{\zeta [ \rho} \, \bar n_{\sigma ]} 
		+ 4 \, \bar n_{[ \mu} \, \dot{\bar{K}}_{\nu ] [ \rho} \bar n_{\sigma ]} \, , 
		\label{eq:3+1_Riemann}
\end{align}
where ${}^3 \! \bar{R}_{\mu \nu \rho \sigma}$ stands for the three-Riemann curvature of the hypersurfaces. 
The over-dot indicates the covariant derivative along the world-lines of the observers 
(for any tensor field $\bm T$, we have: $\dot{T}_{\mu_1 \dots \mu_p} = \bar n^\rho \bar \nabla_\rho T_{\mu_1 \dots \mu_p}$). 
The purely spatial projection of \eqref{eq:3+1_Riemann} only calls upon the first two terms, and it drives the Gauss relation 
\be
	\bar h^\varphi_{\phantom{\alpha} \mu} \bar h^\upsilon_{\phantom{\upsilon} \nu} 
	\bar h^\xi_{\phantom{\xi} \rho} \bar h^\zeta_{\phantom{\zeta} \sigma} \, 
	\bar R_{\varphi \upsilon \xi \zeta} 
		= \, {}^3 \! \bar{R}_{\mu \nu \rho \sigma} + 2 \bar K_{\mu [ \rho} \bar K_{\sigma ] \nu} \, . 
\ee
The three-space and one-time projection gives, from the next two terms, the Codazzi relation 
\be
	\bar h^\varphi_{\phantom{\varphi} \mu} \bar
        h^\upsilon_{\phantom{\upsilon} \nu} \bar
        h^\xi_{\phantom{\xi} \rho} \bar n^\zeta \, 
	\bar R_{\varphi \upsilon \xi \zeta} 
		= \bar D_\mu \bar K_{\nu \rho} - \bar D_\nu \bar K_{\mu \rho} \, , 
	\label{eq:codazzi}
\ee
and the last non-null projection (two-space and two-time) provides, from the last two terms, an evolution equation for the extrinsic curvature. 
These decompositions can be performed in \textit{xAct} with the function {\tt GaussCodazzi}. 

For FLRW space--times, the curvature tensors of the hypersurfaces cast the form 
\be
	{}^3 \! \bar R_{\mu \nu \rho \sigma} 
		= 2 k \, \bar h_{\rho [ \mu} \, \bar h_{\nu] \sigma} \, , \qquad
	{}^3 \! \bar R_{\mu \nu} 
		= 2 k \, \bar h_{\mu \nu} \, , \qquad 
	{}^3 \! \bar R 
		= 6 k \, , 
	\label{eq:curv_tens_FLRW}
\ee
with $k$ being the curvature parameter (equal to zero for flat FLRW cosmologies). 
The corresponding expressions for general Bianchi space--times are more involved, 
as they require the introduction of the constants of structures.
We detail their derivation in appendix~\ref{sec:bianchi_cosmo}. 

%---Subsection
\subsection{Derivatives}

In order to achieve the $3+1$ splitting, we are left with the decomposition of the covariant derivative 
$\bm{\bar \nabla}$ in terms of the induced derivative $\bar{\bm D}$. 
For general spatial tensors (namely, for spatial tensors defined within $\, \overline{\! \cal M}$ or defined within $\cal M$ 
then mapped onto $\, \overline{\! \cal M}$), the relation between the two derivatives reads 
\begin{align}
	\bar \nabla_\rho  T_{\mu_1 \dots \mu_p} 
		= - \bar n_\rho \dot{T}_{\mu_1 \dots \mu_p} 
		+ \bar D_\rho T_{\mu_1 \dots \mu_p} 
		+ \sum_{i=1}^{p} \bar n_{\mu_i} \, \bar K^{\sigma}_{\phantom{\sigma} \rho} \, T_{\mu_1 \dots \mu_{i-1} \sigma \mu_{i+1} \dots \mu_p} \, . 
	\label{eq:cov_der}
\end{align}

Even though our formalism is purely geometrical, we aim at eventually providing, for the perturbations, 
partial differential equations with respect to the proper time $\eta$ of the Eulerian observers. 
When considering the four-dimensional basis built to address the Bianchi classification (refer to appendix~\ref{sec:bianchi_cosmo}), 
the Lie derivative along the direction of $\bm{\bar n}$ precisely comes down to $\partial_\eta$. 
It is accordingly more appropriate for our purpose to use the Lie derivative rather than the dot derivative. 

The relation between $\Lie_{\bm{\bar n}}$ and $\bar n^\rho \bar \nabla_\rho$ is written as\footnote{%
Note that for a spatial tensor $\bm T$, the quantity $\Lie_{\bm{\bar n}} T_{\mu_1 \dots \mu_p}$ is also spatial. 
} 
%
% Expanded expression: 
%\be
%	\Lie_{\bm{\bar n}}  T_{\mu_1 \dots \mu_p} 
%		= \dot{T}_{\mu_1 \dots \mu_p} 
%		\; + \; T_{\alpha \mu_2 \dots \mu_p} \, \bar K^\alpha_{\phantom{\alpha} \mu_1} 
%		\; + \; \dots \; 
%		+ \; T_{\mu_1 \dots \mu_{p-1} \alpha} \, \bar K^\alpha_{\phantom{\alpha} \mu_p} \, . 
%	\label{eq:lie_dot_relat}
%\ee
%
% Contracted expression: 
\be
	\Lie_{\bm{\bar n}}  T_{\mu_1 \dots \mu_p} 
		= \dot{T}_{\mu_1 \dots \mu_p} 
		+ \sum_{i=1}^{p} \bar K^{\sigma}_{\phantom{\sigma} \mu_i} \, T_{\mu_1 \dots \mu_{i-1} \sigma \mu_{i+1} \dots \mu_p} \, , 
	\label{eq:lie_dot_relat}
\ee
and provides us with the following reformulation of \eqref{eq:cov_der}: 
%
% Expanded expression: 
%\be \label{eq:CDTocd}
%	\bar \nabla_\alpha  T_{\mu_1 \dots \mu_p} 
%		= - \bar n_\alpha \, \Lie_{\bm{\bar n}} T_{\mu_1 \dots \mu_p} \, + \, \bar D_\alpha T_{\mu_1 \dots \mu_p} 
%		+ \; 2 \, \bar n_{( \alpha} \, \bar K_{\mu_1 )}^{\phantom{\mu_1} \beta} \, T_{\beta \mu_2 \dots \mu_p} 
%		+ \; \dots \; 
%		+ \; 2 \, \bar n_{( \alpha} \, \bar K_{\mu_p )}^{\phantom{\mu_p} \beta} \, T_{\mu_1 \dots \mu_{p-1} \beta} \, . 
%\ee
%
% Contracted expression: 
\be
	\bar \nabla_\rho  T_{\mu_1 \dots \mu_p} 
		= - \bar n_\rho \, \Lie_{\bm{\bar n}} T_{\mu_1 \dots \mu_p} \, + \, \bar D_\rho T_{\mu_1 \dots \mu_p} 
		+ 2 \sum_{i=1}^{p} \bar n_{(\mu_i} \, \bar K^{\sigma}_{\phantom{\sigma} \rho)} \, T_{\mu_1 \dots \mu_{i-1} \sigma \mu_{i+1} \dots \mu_p} \, . 
	\label{eq:CDTocd}
\ee
Finally, the last expression we shall need is the commutation rule between the derivatives $\Lie_{\bm{\bar n}}$ and $\bar{\bm D}$. 
For general spatial 
%
%one-forms the commutation 
%it reads: 
%\begin{align}
%	\left[ \Lie_{\gr{\bar n}} ,\bar D_\mu \right] \bar \omega_\nu 
%		 = \left( \bar h^{\alpha \beta} \bar D_\beta \bar K_{\mu \nu}  
%		 - \bar D_\mu \bar K_\nu^{\phantom{\nu} \alpha} 
%		 - \bar D_\nu \bar K_\mu^{\phantom{\mu} \alpha} \right) \bar \omega_\alpha \, , 
%\end{align}
%and for spatial
%
tensors, it is given by 
\begin{align} \label{eq:commution_cov_lie}
%	\big[ \Lie_{\bm{\bar n}}, \bar D_\mu \big] \, T_{\nu_1 \dots \nu_p} 
	\Lie_{\bm{\bar n}} \left( \bar D_\rho T_{\mu_1 \dots \mu_p} \right) 
		%
%		& = \Big( \bar h^{\alpha \beta} \bar D_\beta \bar K_{\mu \nu_1} 
%		- \bar D_\mu \bar K_{\nu_1}^{\phantom{\nu_1} \alpha} 
%		- \bar D_{\nu_1} \bar K_\mu^{\phantom{\mu} \alpha} \Big) \, T_{\alpha \nu_2 \dots \nu_p} 
		%  
%		+ \dots 
		% 
%		+ \Big( \bar h^{\alpha \beta} \bar D_\beta \bar K_{\mu \nu_p} 
%		- \bar D_\mu \bar K_{\nu_p}^{\phantom{\nu_p} \alpha} 
%		- \bar D_{\nu_p} \bar K_\mu^{\phantom{\mu} \alpha} \Big) \, T_{\nu_1 \dots \nu_{p-1} \alpha} \, , \nonumber \\ 
		% 
%
% The above commented expression is the expanded one. 
%
		& = 
		\bar D_\rho \left( \Lie_{\bm{\bar n}} T_{\mu_1 \dots \mu_p} \right) + \, 
		\sum_{i = 1}^{p} \big( \, \bar h^{\sigma \zeta} \bar D_\zeta \bar K_{\rho \mu_i} 
		- \bar D_\rho \bar K_{\mu_i}^{\phantom{\mu_i} \sigma} 
		- \bar D_{\mu_i} \bar K_\rho^{\phantom{\rho} \sigma} \, \big) \, T_{\mu_1 \dots \mu_{i-1} \sigma \mu_{i+1} \dots \mu_p} \, , 
\end{align}
where we have made use of relations \eqref{eq:cdtoCD} and \eqref{eq:codazzi} for its derivation. 

%---Subsection
\subsection{Implementation in \xPand}

The $3+1$ splitting of the background manifold is performed by the \textit{xPand} function {\tt SetSlicing}. 
It can be applied to the following spatially homogeneous cosmologies: 
{\tt "Minkowski"}, {\tt "FLFlat"}, {\tt "FLCurved"}, {\tt "BianchiI"}, {\tt "BianchiA"}, {\tt "BianchiB"} and {\tt "Anisotropic"}. 

%_Subsubsection
\subsubsection{Construction of the spatial hypersurfaces}

From the ambient metric {\tt g}, {\tt SetSlicing} first defines the unit normal vector {\tt n}, the induced metric {\tt h} 
of the hypersurfaces, the associated covariant derivative {\tt cd}, the associated scale factor {\tt a[h]} and the 
conformal metric {\tt gah2}. It then specifies the expressions of the intrinsic and extrinsic curvature tensors 
according to the type of cosmologies chosen by the user (hence, the geometry of the model is fully described). 

Let us illustrate with the cosmologies {\tt "FLCurved"} and {\tt "BianchiA"} 
(for the latter most sophisticated case, refer to appendix~\ref{sec:bianchi_cosmo}). 
With the evaluation 
\mathIn{SetSlicing[ g, n, h, cd, \{"|","$\bar{\tt D}$"\}, "FLCurved" ];} 
the extrinsic curvature {\tt K[h][-$\alpha$,-$\beta$]} is set to zero, and the expressions of the 
three-curvature tensors are implemented following relations \eqref{eq:curv_tens_FLRW}. 
For instance, we have for the three-Ricci tensor
\mathIn{Riccicd[-$\alpha$,-$\beta$]}
\mathOut{$2 \, \bar h_{\bb\cc} \, k $}
For {\tt "BianchiA"} cosmologies, 
\mathIn{SetSlicing[ g, nA, hA, cdA, \{"|","$\bar{\tt D}$"\}, "BianchiA" ];}
{\tt SetSlicing} defines the spatial constants of structure ${\cal C}^k{}_{i j}$ (see appendix~\ref{sec:bianchi_cosmo}) 
and then constructs a function allowing one to express the intrinsic curvature tensors in terms of them. 
For the three-Ricci tensor, we then have 
\mathIn{RiccicdA[-b,-c] // ToConstantsOfStructure[hA]}
\mathOut{$ \displaystyle 
	\frac{1}{4} \, {\cal C}_{\bb}^{\phantom{\bb} \lambda \ii} \, {\cal C}_{\cc \lambda \ii} - 
	\frac{1}{2} \, {\cal C}_{\lambda \cc \ii} \, {\cal C}^{\lambda \phantom{\bb} \ii}_{\phantom{\ii} \bb} - 
	\frac{1}{2} \, {\cal C}^{\lambda \phantom{\bb} \ii}_{\phantom{\ii} \bb} \, {\cal C}_{\ii \cc \lambda} + 
	\frac{1}{2} \, {\cal C}_{\bb \cc}^{\phantom{\bb \cc} \lambda} \, {\cal C}^\ii_{\phantom{\jj} \lambda \ii} + 
	\frac{1}{2} \, {\cal C}_{\cc \bb}^{\phantom{\cc \bb} \lambda}
        \, {\cal C}^\ii_{\phantom{\jj} \lambda \ii} + 
	\frac{1}{2} \, {\cal C}^{\lambda}_{\phantom{\ii} \bb \cc} \,
        {\cal C}^\ii_{\phantom{\jj} \lambda \ii} $}
which is equivalent to equation~\eqref{eq:RicciToConstants} \footnote{%
Equation \eqref{eq:RicciToConstants} is recovered by making use of the Jacobi identity \eqref{eq:jacobi} on \textit{Out[\theinout]}. 
This latter relation is not implemented, as it implies a symmetry among several terms that \textit{xTensor} does not yet handle. 
However, once the constants of structure are expanded according to the parameterization~\eqref{eq:ToBianchiType}, 
the Jacobi identity reduces to the constraint \eqref{eq:na}, which is automatically applied in \textit{xPand}.
}. 
The constants of structure can be further expanded, following the usual 
Sch\"ucking, Kundt and Behr (SKB) decomposition \cite{1969CMaPh..12..108E}, 
by means of the \textit{xPand} function {\tt ToBianchiType[hA]}. 

We summarize in table~\ref{tab:eval_curv} the different evaluations 
performed by {\tt SetSlicing} with respect to the type of cosmologies. 

\vspace{2mm}
\begin{table}[htb]
	\begin{tabular}{c|c|c|c}
		\textit{Space--time} & \, \textit{Extrinsic curvature} \, & \textit{Three-curvature tensors} & \textit{Constants of structure} \\[3pt]
		\hline
		&&&\\[-7pt]
		{\tt Minkowski} & Null & Null & Null \\
		&&&\\
		{\tt FLFlat} & Null & Null & Null \\
		&&&\\
		{\tt FLCurved} & Null & Eqs.~\eqref{eq:curv_tens_FLRW} & ${\cal C}^k_{\phantom{k} i j} = 2 \sqrt{k} \; \bar h^{mk} \, \epsilon_{m i j} $ \\ 
		&&&\\
		{\tt BianchiI} & $\bar K_{\mu \nu}$ & Null & Null\\
		&&&\\
		{\tt BianchiA} & $\bar K_{\mu \nu}$ & \, Eqs.~\eqref{eq:RiemannToConstants}, \eqref{eq:RicciToConstants}, \eqref{eq:RicciScalToConstants} \, 
			& ${\cal C}^k_{\phantom{k} i j} = \epsilon_{i j m} N^{m k}$\\ 
		&&&\\
		{\tt BianchiB} & $\bar K_{\mu \nu}$ & Eqs.~\eqref{eq:RiemannToConstants}, \eqref{eq:RicciToConstants}, \eqref{eq:RicciScalToConstants} 
			& \, ${\cal C}^k_{\phantom{k} i j} = \epsilon_{i j m} N^{m k} + 2 A_{[i} \delta^k_{\phantom{k} j]}$\\
		&&&\\
		\, {\tt Anisotropic} \, & $\bar K_{\mu \nu}$ & Eqs.~\eqref{eq:RiemannToConstants}, \eqref{eq:RicciToConstants}, \eqref{eq:RicciScalToConstants} 
			& ${\cal C}^k_{\phantom{k} i j}$\\[3pt]
	\end{tabular}
	\vspace{1mm}
	\caption{Evaluations performed by {\tt SetSlicing} for the extrinsic curvature, the three-curvature tensors and the constants of structure, 
		according to the type of homogeneous cosmologies. In terms of {\tt g} (or, equivalently, $\bm{\bar g}$), the dynamics of 
		{\tt "Minkowski"} and {\tt "FLFlat"} space--times are identical. The difference lies in the conformal transformation \eqref{eq:conf_met_back}: 
		for the former models, the scale factor is set to 1. For {\tt "BianchiA"}, {\tt "BianchiB"} and {\tt "Anisotropic"} space--times, 
		the three-curvature tensors can be formulated in terms of the constants of structure by means of the function {\tt ToConstantsOfStructure[]} 
		and, for the two first models, can be further expanded with the help of {\tt ToBianchiType[]}. The {\tt "Anisotropic"} space--time is used for hypersurfaces 
		whose dimension differs from three (for these models the SKB decomposition does not apply).} 
	\label{tab:eval_curv}
\end{table}

%_Subsubsection
\subsubsection{Derivatives}

For anisotropic cosmologies, {\tt SetSlicing} creates a set of automatic rules to handle the action 
of the derivatives on the three-curvature tensors. The aim is to obtain a formulation in terms of the 
extrinsic curvature tensor and/or the constants of structure, which fully identify the geometry of the model. 
Note that such rules are not necessary for isotropic models: the expressions of the intrinsic curvature 
tensors are simple enough to let \textit{xTensor} perform the evaluation. 

The action of the derivative $\bm{\bar \nabla}$ is implemented following equation \eqref{eq:CDTocd}, 
that of $\bar{\bm D}$ following equation \eqref{eq:cdOfBackgd} together with \eqref{eq:GammaToC}, 
and the application of $\Lie_{\bm{\bar n}}$ on a three-curvature tensor automatically calls the 
function {\tt ToConstantsOfStructure[]} so as to use the property~\eqref{eq:LieC}. 
Regarding the three-covariant derivative, we have for instance 
\mathIn{cdA[-$\bb$]@RiccicdA[-$\ii$,-$\jj$]}
\mathOutTwoLines{$ \displaystyle 
	\frac{1}{2} \, {\cal C}_{\bb \jj}^{\phantom{\bb \jj} \cc} \, R[\bar D]_{\ii \cc} - 
	\frac{1}{2} \, {\cal C}^\cc_{\phantom{\cc} \bb \jj} \, R[\bar D]_{\ii \cc} + 
	\frac{1}{2} \, {\cal C}_{\jj \bb}^{\phantom{\jj \bb} \cc} \, R[\bar D]_{\ii \cc} + 
	\frac{1}{2} \, {\cal C}_{\bb \ii}^{\phantom{\bb \ii} \cc} \, R[\bar D]_{\jj \cc} -
	\frac{1}{2} \, {\cal C}^\cc_{\phantom{\cc} \bb \ii} \, R[\bar D]_{\jj \cc}$} 
	{$ \displaystyle + 
	\frac{1}{2} \, {\cal C}_{\ii \bb}^{\phantom{\ii \bb} \cc} \, R[\bar D]_{\jj \cc} $}

Lastly, {\tt SetSlicing} constructs an automatic rule to perform the commutation \eqref{eq:commution_cov_lie} 
for \textit{any} expression. This serves to make sure that the Lie derivative will first act on a given tensor, 
so as to recover the usual formulation of the perturbed equations\footnote{%
In line of this comment, let us note that {\tt SetSlicing} also creates \textit{internal} rules for the commutation 
of several $\bar{\bm D}$. These will enforce the appearance of Laplacians, and move a covariant derivative 
closer to a tensor when a divergence is present.
}. For {\tt "FLCurved"} cosmologies, we have 
\mathIn{DefTensor[ V[-$\bb$], M, OrthogonalTo->\{n[$\bb$]\}, ProjectedWith->\{h[$\bb$,-$\cc$]\} ];}
\mathIn{LieD[n[$\ii$]]@cd[-$\jj$]@V[-$\bb$]}
\mathOut{$\bar D_\jj \, \Lie_{\bar n} \, V_\bb$}
and for {\tt "BianchiA"}, 
\mathIn{DefTensor[TA[-$\bb$]\!,M,OrthogonalTo->\{nA[$\bb$]\},ProjectedWith->\{hA[$\bb$,-$\cc$]\},PrintAs->"T"];}
\mathIn{LieD[nA[$\ii$]]@cdA[-$\jj$]@TA[-$\bb$]}
\mathOut{${\cal C}^{\lambda \phantom{\jj} \cc}_{\phantom{\ii} \jj} \, K_{\bb \lambda} \, T_\cc 
	- {\cal C}_{\bb \jj}^{\phantom{\bb \jj} \lambda} \, K^\cc_{\phantom{\cc} \lambda} \, T_\cc 
	- {\cal C}_{\jj \bb}^{\phantom{\jj \bb} \lambda} \, K^\cc_{\phantom{\cc} \lambda} \, T_\cc 
	+ {\cal C}^{\lambda \phantom{\bb} \cc}_{\phantom{\ii} \bb} \, K_{\jj \lambda} \, T_\cc 
	+ \bar D_\jj \, {\cal L}_{\bar{nA}} \, T_\bb $}
where to obtain \textit{Out[\theinout]}, expression \eqref{eq:cdOfBackgd} has been automatically applied 
to the extrinsic curvature tensor.

%%% Section
\section{Perturbed fields} \label{sec:pert_fields}

We now present the splitting of the perturbed fields into their spatial and temporal parts, 
the scalar--vector--tensor (SVT) decomposition of their projected components and finally the parameterization 
of the metric perturbations.

% In this section we detail how the perturbation of the tensors, which
% according to \S\ref{secGeneral} live on the background space--time, are
% decomposed according to this background $1+3$ splitting.

%---Subsection
\subsection{Projected components and scalar-vector-tensor decomposition}

Any perturbed quantity lying within $\cal M$ and mapped onto the background manifold can be decomposed 
by means of the normal vector $\bm{\bar n}$ and the induced metric $\bm{\bar h}$. 
For a rank-2 covariant tensor $\bm T$, we have for instance, 
\be
	T_{\mu \nu} 
		= \bar n_\mu \bar n_\nu \left( \, \bar n^\rho \bar n^\sigma T_{\rho \sigma} \right) 
		+ 2 \, \bar n_{( \mu} \big( \, \bar n^\rho \, \bar h^{\sigma}_{\phantom{\sigma} \nu )} T_{\rho \sigma} \big) 
		+ \big( \, \bar h^{\rho}_{\phantom{\rho} \mu} \bar h^{\sigma}_{\phantom{\sigma} \nu} T_{\rho \sigma} \big) \, .
\ee
The terms inside parentheses hence constructed respectively define a scalar, a spatial vector and a spatial tensor. 
As is customary in perturbation theory, we shall then employ the SVT decomposition to 
further expand the two last projected quantities \cite{Bardeen:1980kt}. 
Let us briefly recall the form of this decomposition for later needs: 
a spatial vector $U_\mu$ is split into a scalar part $S$ and a vector part $V_\mu$ as 
\be
	U_\mu = \bar D_\mu S + V_\mu \, , \qquad 
		\mathrm{with} \quad 
	\bar D^\mu V_\mu = 0 \, ,
\ee
and a symmetric spatial tensor $H_{\mu\nu}$ is decomposed into two
scalar parts $S_1$ and $S_2$, a vector part $V_\mu$ 
and a tensor part $T_{\mu \nu}$ as 
\be
	H_{\mu\nu} 
		= \bar D_\mu \bar D_\nu S_1 + S_2 \bar h_{\mu\nu}  
		+ \bar D_{(\mu} V_{\nu)} + T_{\mu\nu} \, , \qquad 
	\mathrm{with} \quad
	\left\{
		\begin{array}{l}
			\bar D^\mu V_\mu = 0 \, , \\ 
			\bar D^\mu T_{\mu\nu} = 0 \quad \mathrm{and} \quad T^\mu_{\phantom{\mu} \mu} = 0 \, . 
		\end{array}
	\right.
\ee

%---Subsection
\subsection{Perturbations of the metric} \label{subsec:pert_met}

The SVT decomposition of the metric perturbations yields the general expressions\footnote{%
Note that for Bianchi space--times it is more convenient to modify this decomposition by replacing the last term of equation~\eqref{eq:ss_pert_met} with 
$\pert{n}{\psi} 3 \bar K_{\mu \nu} / \bar K^\alpha_\alpha$ \cite{Pitrou:2008gk}. This boils down to a redefinition of $\pert{n}{E}$, $\pert{n}{E_\mu}$,
$\pert{n}{E_{\mu \nu}}$ and $\pert{n}{\psi}$, and it is this parameterization that we have chosen in \textit{xPand}.
} 
\begin{align}
	\bar n^\rho \bar n^\sigma \pert{n}{h_{\rho \sigma}} 
		& = -2 \pert{n}{\phi} \, , 
		\label{eq:tt_pert_met} \\ 
	\bar n^\rho \bar h^{\sigma}_{\phantom{\sigma} \nu} \! \pert{n}{h_{\rho \sigma}} 
		& = - \bar D_\nu \pert{n}{B} - \pert{n}{B_\nu} \, , 
		\label{eq:ts_pert_met} \\
	\bar h^{\rho}_{\phantom{\rho} \mu} \bar h^{\sigma}_{\phantom{\sigma} \nu} \pert{n}{h_{\rho \sigma}} 
		& = 2 \left( \bar D_\mu \bar D_\nu \pert{n}{E} + \bar D_{(\mu} \pert{n}{E_{\nu)}} 
		+ \pert{n}{E_{\mu\nu}} - \pert{n}{\psi} \, \bar h_{\mu\nu} \right) \, .  
		\label{eq:ss_pert_met}
\end{align}
Four of the ten degrees of freedom of $\! \pert{n}{\bm h}$ are carried by the scalars 
$\! \pert{n}{\phi}$, $\! \pert{n}{\psi}$, $\! \pert{n}{E}$ and $\! \pert{n}{B}$,
four are encoded in the vectors $\! \pert{n}{E_\mu}$ and $\! \pert{n}{B_\nu}$, 
and two are contained in the tensor $\! \pert{n}{E_{\mu \nu}}$.
Note that some of these fields are required to vanish for specific gauge choices 
(see, e.g., \cite{Malik:2008im} for a comprehensive review). 

%---Subsection
\subsection{Implementation in \xPand}

%_Subsubsection
\subsubsection{Construction of SVT quantities}

The construction of spatial tensors satisfying the SVT properties is performed 
by the \textit{xPand} function {\tt Def\-Projected\-Tensor}. By default, these tensors 
are defined on both the background and perturbed manifolds. 
For instance, we have 
\mathIn{DefProjectedTensor[ U[-$\bb$,-$\cc$], h ];}
\mathIn{n[-$\bb$] U[$\bb$, $\cc$]}
\mathOut{$0$}
\mathIn{U[$\bb$, -$\bb$]}
\mathOut{$0$}
\mathIn{Perturbation[ U[-$\bb$,-$\cc$], 1 ]}
\mathOut{$\Delta \big[ \, U_{\bb\cc} \, \big]$}
To relax one or several default properties, the user has to modify the optional arguments 
{\tt TensorProperties}, set by default to {\tt \{"SymmetricTensor","Traceless","Transverse"\}}, 
and {\tt SpaceTimesOfDefinition}, evaluated by default as {\tt \{"Background","Perturbed"\}}: 
\mathIn{UndefTensor[ U ];}
\mathInTwoLines{DefProjectedTensor[\,U[-$\bb$,-$\cc$]\!,\,h,\,TensorProperties->\{"SymmetricTensor","Transverse"\},}%
		{\,SpaceTimesOfDefinition->\{"Background"\}];}
\mathIn{U[$\bb$,-$\bb$]}
\mathOut{$U^{\bb}_{\phantom{\bb} \bb}$}
\mathIn{Perturbation[U[-$\bb$, -$\cc$], 1]}
\mathOut{$0$}
%

%_Subsubsection
\subsubsection{Comments on the label-indices}

In the internal notation, \textit{xPert} attaches a label-index to the metric perturbations in order to denote 
their order. The first order perturbation $dg^{1}_{\alpha \beta}$ (cf output \textit{Out[5]}) is hence stored as {\tt dg[LI[1],-$\alpha$,-$\beta$]}. 
In \textit{xPand}, we employ the same notation for spatial tensors defined with {\tt DefProjectedTensor}, 
and we moreover attach a second label-index to indicate the number of Lie derivatives along $\bm{\bar n}$ acting on them. 
Thus, we have for instance 
\mathIn{DefProjectedTensor[V[-$\bb$], h]}
\mathIn{LieD[n[$\ii$]]@V[LI[1], LI[0], -$\bb$]}
\mathOut{${}^{(1)} {V_\bb}'$}
\mathIn{V[LI[1], LI[1], -$\bb$]}
\mathOut{${}^{(1)} {V_\bb}'$}
{\tt DefProjectedTensor} constructs a set of rules to automatically allocate label-indices to a tensor written without. 
Hence, {\tt V[-$\bb$]} is converted to {\tt V[LI[0],LI[0],-$\bb$]}, while {\tt V[LI[p],-$\bb$]} is converted into 
{\tt V[LI[p],LI[0],-$\bb$]}, for any perturbation order {\tt p}. 

Note that \textit{xTensor} interprets a tensor with label-indices as a tensor by itself, whatever the meaning of the label-indices. 
Its indices are therefore raised and lowered by the ambient metric in the usual way: 
{\tt g[$\alpha$,$\beta$] V[LI[1],LI[1],-$\bb$]} yields {\tt V[LI[1],LI[1],$\cc$]}. 
While this is obviously mathematically correct for conformal isotropic manifolds\footnote{%
In such cases, {\tt g[$\alpha$,$\beta$] \!\!\!\!\! V[LI[1],LI[1],-$\alpha$]} corresponds to $g^{\alpha \beta} \Lie_{\bm{\bar n}} V_\alpha$ 
which is equal to $\Lie_{\bm{\bar n}} V^\beta$, as the extrinsic curvature vanishes. This last expression indeed corresponds to {\tt V[LI[1],LI[1],$\beta$]}. 
}, it is no longer true for anisotropic ones. For these latter models, the second label-index can be interpreted 
as Lie derivatives \textit{only} when the tensor is with indices \textit{down}. To avoid confusion, 
we modify the output when the tensor indices are up. For instance, we have for {\tt "BianchiA"} cosmologies 
\mathIn{DefProjectedTensor[VA[-$\bb$], hA]}
\mathIn{VA[LI[1], LI[1], -$\bb$]}
\mathOut{${}^{(1)} {V_\bb}'$}
\mathIn{VA[LI[1], LI[1], $\bb$]}
\mathOut{${}^{(1)} \overset{1}{V^\bb}$}
%

%_Subsubsection
\subsubsection{Construction of the perturbations of the metric}\label{secConspert}

The perturbed fields introduced in subsection \ref{subsec:pert_met} are constructed with the 
\textit{xPand} function {\tt DefMetricFields}. The evaluation of the command 
\mathIn{DefMetricFields[ g, dg, h ];}
calls the function {\tt DefProjectedTensor} in order to define the projected components of the metric perturbations 
and allocate them all the SVT properties. The set of rules \eqref{eq:tt_pert_met}--\eqref{eq:ss_pert_met} 
are then automatically defined using the \textit{xPand} function {\tt SplitMetric}: 
\mathIn{GaugeRules\ =\ SplitMetric[ g, dg, h, "AnyGauge" ]}
\mathIn{dg[LI[1], -$\bb$, -$\cc$]  /.\ GaugeRules }
\mathOutTwoLines{$2 \, {}^{(1)} \! E_{\bb\cc} 
	+ {}^{(1)} \! B_{\cc} \, \bar n_\bb 
	- {}^{(1)} \! B_\bb \, \bar n_\cc 
	- 2 \, \bar n_\bb \, \bar n_\cc \, {}^{(1)} \! \phi 
	- 2 \, \bar h_{\bb\cc} \, {}^{(1)} \! \psi 
	- \bar n_\cc \, \bar D_\bb {}^{(1)} \! B 
	+ \bar D_\bb \, {}^{(1)} \! E_\cc 
	- \bar n_\bb \, \bar D_\cc \, {}^{(1)} \! B $}%
	{$ + \bar D_\cc \, {}^{(1)} \! E_\bb
	+ 2 \, \bar D_\cc \, \bar D_\bb \, {}^{(1)} \! E $}

The particular gauges we have implemented are: 
{\tt "ComovingGauge"}, {\tt "FlatGauge"}, {\tt "IsoDensityGauge"}, {\tt "Newton\-Gauge"} and {\tt "SynchronousGauge"}. 
Should the user wish to consider other gauges, own rules can be created as follows: 
{\tt MyGauge = \{dg[LI[ord\_], $\ii$\_ ,$\jj$\_] :> ...\}}. 

%_Subsubsection
\subsubsection{Splitting of the background covariant derivative} \label{1p3perturbed}

We now have nearly all the necessary tools to obtain the final expression of any perturbed field. 
Let us quickly review the previous steps and then introduce the last function we need. 

To derive the perturbation of, e.g., the four-dimensional Ricci scalar, we first perform a conformal transformation from 
{\tt g} to {\tt gah2} (and express it in terms of the metric {\tt g}), then we perturb the resulting expression, and finally 
we substitute the metric perturbations by their SVT components using the set of rules {\tt GaugeRules}: 
\mathIn{Conformal[g, gah2][ RicciScalarCD[] ]}
\mathIn{MyR = ExpandPerturbation@Perturbed[ \%, 1 ]}
\mathInTwoLines{(MyR\ /.GaugeRules)\ // ProjectorToMetric\ // GradNormalToExtrinsicK}{ // ContractMetric\ // ToCanonical}
The final result has been expanded using the \textit{xTensor} functions {\tt ProjectorToMetric}, 
which replaces $\bar h_{\mu\nu}$ by $\bar g_{\mu\nu} - \bar n_\mu \bar n_\nu$, 
and {\tt GradNormalToExtrinsicK}, which replaces $\bar{\nabla}_\mu \bar n_\nu$ by $\bar{K}_{\mu\nu}$, 
and it has been simplified with {\tt ContractMetric} and {\tt ToCanonical}. 

Since the relation \eqref{eq:CDTocd} is not automatically evaluated so far for quantities other than the three-curvature tensors, 
the result of \textit{In[\theinout]} still involves the covariant derivative $\bm{\bar \nabla}$. In order to split the latter in terms 
of the induced derivative and Lie derivative along $\bm{\bar n}$, one finally needs to use the \textit{xPand} function {\tt SplitPerturbations}, 
which applies as well the Gauss--Codazzi decompositions. 
In the above example, we obtain 
\mathIn{SplitPerturbations[\% ah[]\^{}2 , h]}
\mathOutTwoLines{$%
	6 \, {\cal H}^2 + 6 \, {\cal H}' + 6 \, \mathcal{K} + 
	\varepsilon \left(-12 \, {\cal H}^2 \pertbis{1}{\phi} - 12 \, {\cal H}' \pertbis{1}{\phi} - 
	6 \, {\cal H} \pertbis{1}{\phi'} + 12 \pertbis{1}{\psi} \, {\cal K} - 18 \, {\cal H} \pertbis{1}{\psi'} - 
	6 \, \pertbis{1}{\psi''}\right.$}%
	{$\left. + 6 \, {\cal H} \, D_\bb D^\bb \pertbis{1}{B} - 2 \, D_\bb D^\bb \pertbis{1}{B'} + 
	6 \, {\cal H} \, D_\bb D^\bb \!\pertbis{1}{E'} + 2 \, D_\bb D^\bb \pertbis{1}{E''} - 
	2 \, D_\bb D^\bb \pertbis{1}{\phi} + 
	4 \, D_\bb D^\bb \!\pertbis{1}{\psi} \right)$}
This way of proceeding is, however, rather inefficient for general gauges at higher order. 
Instead, the set of rules {\tt GaugeRules} can be used in an optimized manner by the function {\tt SplitPerturbations} itself: 
\mathIn{ SplitPerturbations[ah[]${}^2$  MyR,\ GaugeRules,\ h]}
In such a way, the rule following relation \eqref{eq:CDTocd} is evaluated on the projected components of the metric 
\textit{before} specifying their SVT decomposition, which takes much less time at higher orders.

%%% Section
\section{Features of the algorithm} \label{sec:summarize}

%---Subsection
\subsection{Summary}

Let us review the main steps that need to be followed in order to derive the perturbation of any fields. 

\begin{enumerate}
	\item[(i)] We define the background manifold {\tt M} (with {\tt DefManifold}) and the ambient metric {\tt g} (with {\tt DefMetric}). 
		The splitting of {\tt M} is realized with {\tt SetSlicing}, according to the type of cosmology chosen by the user. 
	\item[(ii)] We apply to the quantity to be perturbed a conformal transformation from $\bm{\bar g}$ to $\bm{\widetilde{\bar g}}$ 
		and express the result with respect to the quantities defined on {\tt M}. This is done with the function {\tt Conformal}. 
	\item[(iii)] We use the \textit{xPert} tools {\tt Perturbed} and {\tt ExpandPerturbation} to perturb the previous expression at any order, 
		in terms of the metric perturbations {\tt dg} and other tensors. 
	\item[(iv)] We use the functions {\tt DefMetricFields} and {\tt SplitMetric} to define and construct the SVT parameterization of the 
		metric perturbations (thanks to the function {\tt DefTensorProjected}), according to the gauge chosen by the user. 
		The perturbations of the fluid quantities are defined and decomposed with the functions {\tt DefMatterFields} and {\tt SplitMatter} 
		(see further below). 
	\item[(v)] We finally use the function {\tt  SplitPerturbations} to decompose the covariant derivative $\bm{\bar \nabla}$ in terms of 
		the induced derivative and the Lie derivative along $\bm{\bar n}$. The Gauss--Codazzi relations are also applied, and the 
		constructed rules are used to transform the resulting expression. 
\end{enumerate}

These four last steps have been coded all at once in a single function called {\tt ToxPand}. 
We present it further below, after introducing a minimal example. 

%---Subsection
\subsection{A minimal example} \label{subsec:min_ex}

We now propose a brief and self-contained example to illustrate our package. 
%(in appendix~\ref{app:non-trivial}, we provide a non-trivial example in order to present some other features of \textit{xPand}). 
The parameterization of the metric perturbations is here constructed by hand, and we only consider the Bardeen potentials. 

\vspace{3mm}

%\begin{verbatim}
%
{\tt <<xAct/xPand.m;} \\
\indent{\tt DefManifold[ M, 4, \{$\alpha$, $\beta$, $\mu$, $\nu$\} ];} \\
\indent{\tt DefMetric[ -1, g[-$\alpha$,-$\beta$], CD, \{";", "$\bar\nabla$"\} ];} \\
\indent{\tt DefMetricPerturbation[ g, dg, $\epsilon$ ];} \\
\indent{\tt SetSlicing[ g, n, h, cd, \{"|", "D"\}, "FLCurved" ];} \\
\indent{\tt order = 1;} \\
\indent{\tt DefProjectedTensor[ $\phi$[], h ];} \\
\indent{\tt DefProjectedTensor[ $\psi$[], h ];} \\
\indent{\tt MyRicciScalar = ExpandPerturbation@Perturbed[ Conformal[g, gah2][RicciScalarCD[]], order ];}\\
\indent{\tt MyGauge = {dg[LI[ord\_], $\bb$\_, $\cc$\_] :> - 2 n[$\bb$]n[$\cc$] $\phi$[LI[ord]] - 2 h[$\bb$, $\cc$] $\psi$[LI[ord]]};}\\
\indent{\tt SplitPerturbations[ ah[]\^{}2 MyRicciScalar, MyGauge, h ]}
%
%\end{verbatim}

\vspace{3mm}

\noindent
The output generated by the last line is (compare with \textit{Out[46]}) 
\mathOutTwoLines{$%
	6 \, {\cal H}^2 + 6 \, {\cal H}' + 6 \, \mathcal{K} + 
	\varepsilon \left(-12 \, {\cal H}^2 \pertbis{1}{\phi} - 12 \, {\cal H}' \pertbis{1}{\phi} - 
	6 \, {\cal H} \pertbis{1}{\phi'} + 12 \, \pertbis{1}{\psi} {\cal K} - 18 \, {\cal H} \pertbis{1}{\psi'} 
	- 6 \, \pertbis{1}{\psi''}\right.$}%
	{$\left. - 2 \, D_\bb D^\bb \pertbis{1}{\phi} + 4 \, D_\bb D^\bb \pertbis{1}{\psi} \right)$} 

%---Subsection
\subsection{Secondary functions}

Even though the user is free to parameterize tensor perturbations by creating its own rules 
with projected tensors defined from {\tt DefProjectedTensor}, we have seen in paragraph 
\ref{secConspert} that the functions {\tt DefMetricFields} and {\tt SplitMetric} 
can take care of this procedure for the metric perturbations. Similarly, we have implemented 
in \textit{xPand} the functions {\tt DefMatterFields} and {\tt SplitMatter} for 
the definition and parameterization of the fluid perturbations, and more precisely for 
those of the energy density, pressure and fluid 4-velocity.  
%
% Among these, the less obvious parameterization is the
% one of the fluid velocity, since the time component (that is the
% component along $\bar{\gr{n}}$) is determined from the projected part
% of the perturbation, thanks to the normalization condition. 

We also wish to mention that we have extended in \textit{xPand} the \textit{xPert} function 
{\tt GaugeChange} which performs gauge transformations at any order for a given expression. 
Our extension {\tt SplitGaugeChange} executes a $3+1$ splitting of these transformation rules. 
For a glance at the gauge transformations for metric perturbations and the construction of
gauge-invariant variables, see~\cite{Bardeen:1980kt,Bruni:1996im,Malik:2001rm,Nakamura2007}. 
For examples of use of the secondary functions 
%{\tt SplitMetric}, 
{\tt SplitMatter} and {\tt SplitGaugeChange}, we invite the reader to go through the example 
notebooks which are distributed along with the package \textit{xPand}. 

Furthermore, for the cases where one wants to consider the predefined gauges, it is enough, simple and straightforward 
to use the \textit{xPand} function {\tt ToxPand} to obtain the perturbation of any expression. 
For instance, the following five lines suffice to derive that of the four-Ricci scalar in any gauge and up to order 2 

\vspace{3mm}
 {\tt<<xAct/xPand.m;} \\
\indent{\tt  DefManifold[ M, 4, \{$\bb$, $\cc$, $\ii$, $\jj$\} ];} \\
\indent{\tt DefMetric[ -1, g[-$\alpha$, -$\beta$], CD, \{";", "$\bar \nabla$"\} ];} \\
\indent{\tt SetSlicing[ g, n, h, cd, \{"|", "D"\}, "FLCurved" ];} \\
\indent{\tt ToxPand[ RicciScalarCD[], dg, u, du, h, "AnyGauge", 2 ]}
\vspace{3mm}

\noindent
where {\tt u} and {\tt du} are the fluid four-velocity and its perturbation, respectively. 
The function {\tt ToxPand} combines several functions so that the user can easily obtain the desired perturbations
in a given gauge without having to deal with any detail of the algorithm. 

Finally, the \textit{xPand} function {\tt ExtractComponents} allows to extract the projected 
components of any tensor. For a given background slicing, the user only needs to specify 
the type of projection (see appendix~\ref{app:non-trivial} for an example). Note also 
that for a rank-1 (resp.\ rank-2) tensor, the \textit{xPand} function {\tt VisualizeTensor} 
allows us to display all projected components in a vector (resp.\ matrix) form. Again, we refer 
to the example notebooks which are distributed along with \textit{xPand} \cite{xPand}, for more details 
about the syntax of these functions. 

%---Subsection
\subsection{Recovering standard results}

We have checked that with our implementation we recover the standard results of cosmological perturbation theory. 
More precisely, our algorithm is in accordance with 
\begin{itemize}
	\item all first-order results of Einstein equation and stress--energy tensor conservation equation, 
		for flat and curved FLRW space--times in any gauge (see for instance~\cite{Malik:2008im}); 
	\item all first-order results for Bianchi type I background, in the gauge chosen in~\cite{Pereira:2007yy,Dulaney:2010sq}; 
	\item all second-order results of Einstein equation and stress-energy tensor conservation equation, 
		for flat and curved FLRW space--times in the Newtonian gauge (see for instance~\cite{Nakamura2007} for the complete set of equations). 
\end{itemize}
Our package now enables to extend these results to higher order and in any gauge. 
It also allows for the study of any type of perturbed Bianchi cosmologies. 

It is worth mentioning that in order to obtain useful standard
differential equations with respect to the conformal time $\eta$, it is necessary to perform a mode expansion on the hypersurfaces, that is one needs to find the eigenmodes of the spatial Laplacian $\bar D_\mu \bar D^\mu$. 
This is simple for flat FLRW cosmologies, where one just has to employ a Fourier transformation, and it is also well-known for 
curved FLRW models, where hyper-spherical Bessel functions need to be used~\cite{Abbott:1986ct}. However, the procedure 
is still unknown for general Bianchi cosmologies. Apart from the special case of Bianchi type I, where the modes can also be found 
from a Fourier transformation (and thus lead to a simple set of equations~\cite{Pitrou:2008gk}), there is no \textit{general} technique 
to obtain the eigenmodes of the Laplacian for all other types. Only in special cases (see for instance~\cite{Pereira:2012ma}) this has been done explicitly. 

%---Subsection
\subsection{Timings}

\begin{figure}[!htb]
	\includegraphics[width=8cm]{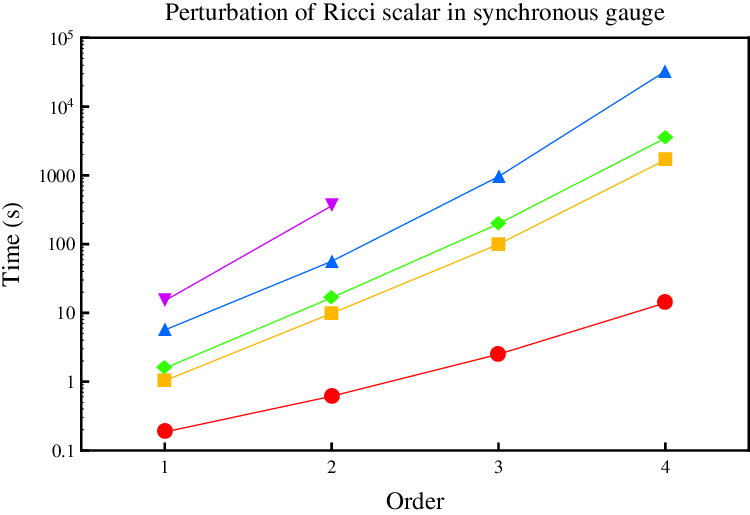}%{Timing.pdf}
	\quad 
	\includegraphics[width=8cm]{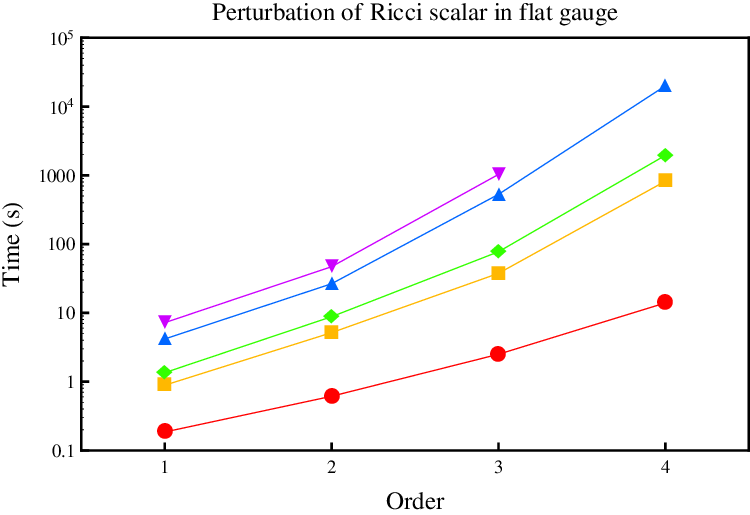} \\%{Timing.pdf}
	\vspace{1mm}
	\includegraphics[width=8cm]{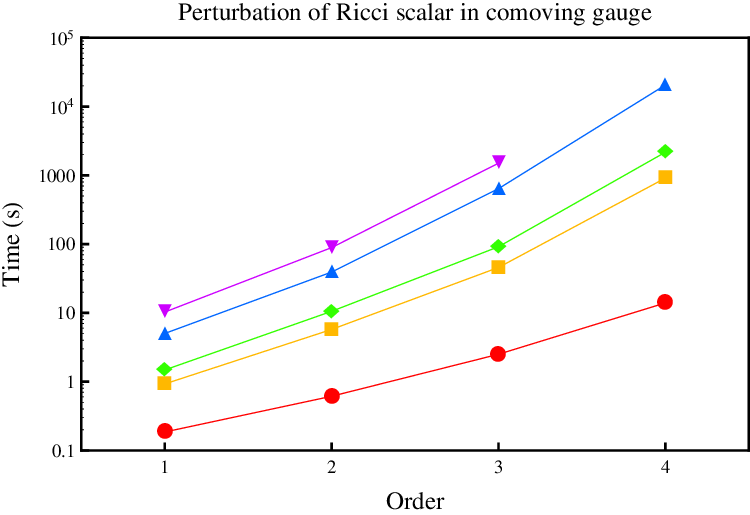}%{Timing.pdf}
	\quad 
	\includegraphics[width=8cm]{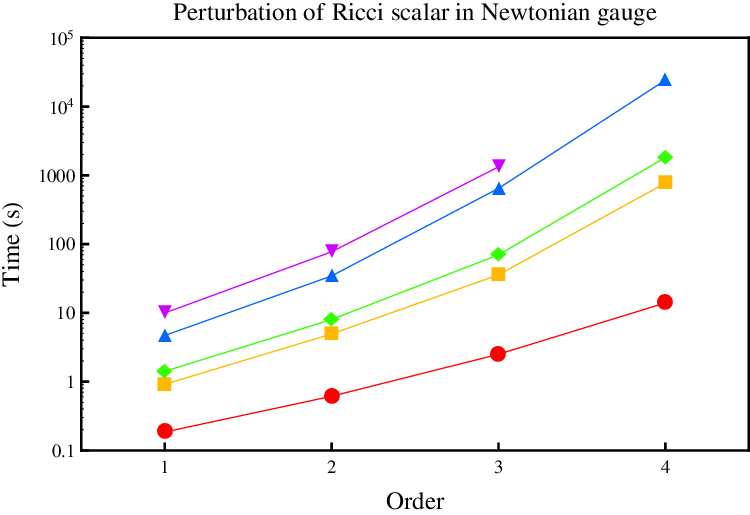}%{Timing.pdf}
	\vspace{-2mm}
	\caption{Timings  for the perturbation of the four-dimensional Ricci scalar in different gauges. 
		From top to bottom and left to right: synchronous gauge, spatially flat gauge, comoving gauge and Newtonian gauge. 
		On each plot, the curves from bottom to top refer to: 
		(i) formal perturbations with \textit{xPert} using a conformal transformation (red line); 
		(ii) perturbations for a Minkowski background (yellow); 
		(iii) perturbations for a curved FLRW background (green); 
		(iv) perturbations for a Bianchi I background (blue); 
		and (v) perturbations for a general Bianchi background ({\tt "BianchiB"}) (purple). 
		All timings were performed on a single $4\,\mathrm{GHz}$ core, with a $8 \,\mathrm{GB}$ RAM.}
	\label{fig1}
	\vspace{-2mm}
\end{figure}

\begin{figure}[!htb]
	\includegraphics[width=8cm]{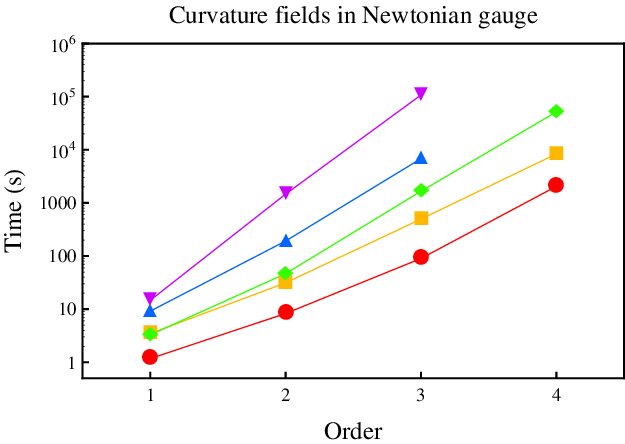}
	\quad 
	\includegraphics[width=8cm]{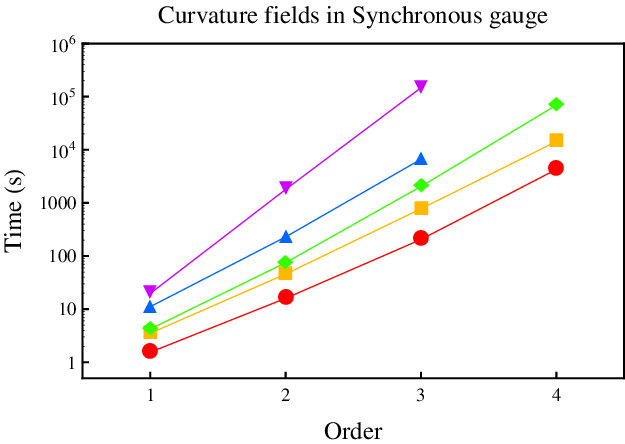}
	\vspace{-2mm}
	\caption{Timings for the perturbations of the four-curvature tensors, for a curved FLRW space--time, in different gauges. 
		Left: Newtonian gauge; right: synchronous gauge.
		On each plot, the curves from bottom to top refer to: 
		(i) Ricci scalar (red line); 
		(ii) Ricci tensor (yellow); 
		(iii) Einstein tensor (green);
		(iv) Riemann tensor (blue); 
		and (v) Weyl tensor (purple). 
		All timings were performed on a single $3.40\,\mathrm{GHz}$ core, with a $8 \,\mathrm{GB}$ RAM.}
	\label{fig2}
\end{figure}

%%%%%%%%%%%%%%%%%%%%%%%%%%%%%%%%%%%%%%%%%%%%%%%%%%%%%%%%%%
In practice, the timing for the computation grows like (slightly faster than) a power law of the perturbation order, 
whatever the gauge (see figure~\ref{fig1}). 
%%%%%%%%%%%%%%%%%%%%% Gauge Restrictions %%%%%%%%%%%%%%%%%%%%%%%%%%%
%%%%%%%%%%%%%%%%%%%%%%%%%%%%%%%%%%%%%%%%%%%%%%%%%%%%%%%%%%
%
It takes \textit{xPand} less than 2 minutes (see figure~\ref{fig2}) to decompose completely the perturbation 
of a rank-2 curvature tensor, such as the Ricci or Einstein tensor, up to second order in any gauge. 
The decomposition for the perturbations of the Riemann or Weyl tensor, up to second order and in any gauge, 
takes a little more than 2 and 13 minutes, respectively.
% depending on the processing power of the  machine. 
%%%%%%%%%%%%%%%%%%%%%%%%%%%%%%%%%%%%%%%%%%%%%%%%%%%%%%%%%%

%%% Section
\section*{Conclusion} \label{sec:conclusion}

\textit{xPand} is the first comprehensive package that allows to perform algebraic calculations in cosmological 
perturbation theory for homogeneous background space--times, up to any
order and in any gauge. 
% It does this in a manner familiar to cosmologists. 
It is worth stressing again the several features and advantages of this package. 

\begin{itemize}
	\item The expression of any perturbed field can be obtained
          for all Bianchi cosmologies, up to any order of interest and in any gauge, in a very simple and straightforward way. 
	\item The package avoids the complexities of a component-by-component computation of the perturbed 
		fields, thanks to the use of the $3+1$ formalism. 
	\item The package is relatively fast: at first order, all the perturbed equations can be derived in approximately 
		two seconds, and as the order of the perturbation increases, the timing grows roughly as a power law. 
	\item The computations can be applied to space--times of arbitrary dimension, and for any extension of general relativity
		(the current exception being any gravitational theory with torsion). 
	\item Taking advantage of the package \textit{xTensor}, \textit{xPand} handles tensor indices just the same 
		way a user would do when calculating by hand. In particular, it does not break the summation over 
		repeated indices (as e.g.\ in \textit{Out[46]}). 
	\item It totally eliminates the laborious summation over repeated indices, an element that has inhibited the 
		use of other packages developed to solve similar models. 
%	\item The user has the freedom to choose a style of formatting/printing for any declared variable. 
	\item The user who has little knowledge of the \textit{Mathematica} or \textit{xTensor} syntax can obtain 
		from the package almost the same utility as an expert would do.
\end{itemize}

Finally, we plan to extend the scope of \textit{xPand} beyond the derivation of perturbed equations in general relativity, 
so as to provide the entire Einstein--Boltzmann system for radiation transfer and the Einstein--Jacobi map system needed 
for understanding the effect of weak gravitational lensing.

%%% Acknowledgements
\acknowledgements

It is a pleasure to thank Jos\'e M. Mart\'in-Garc\'ia and Guillaume Faye for long and fruitful discussions about 
abstract tensor calculus in \textit{xAct}. We also thank Chris Clarkson for comments on the SVT decomposition, 
and Jos\'e M. Mart\'in-Garc\'ia for comments on a previous version of this paper. 
XR acknowledges support from the Claude Leon Foundation. He is grateful for the hospitality received in the 
cosmology group at \textit{`Institut d'Astrophysique de Paris'}, and thanks Jean-Philippe Uzan for his invitation. 
OU is funded by the South African Square Kilometer Array (SKA) Project and Faculty of Science, University of Cape Town. 
This work was supported by French state funds managed by the ANR within the Investissements d'Avenir programme under reference ANR-11-IDEX-0004-02.

%%% Bibliography

% \bibliographystyle{utphys} % {revtex} {Science} {h-physrev5} {JHEP-2}

%

%%% Appendix
\appendix

\setcounter{inout}{0}

%%% Section
\section{What {\xPand} can do: a non-trivial example} \label{app:non-trivial}

We here described by way of an example how \textit{xPand} can be used to derive 
relativistic perturbed equations for flat FLRW backgrounds, up to first order and in 
the Newtonian gauge. An extension to any other background, order or gauge simply 
requires a change of the related arguments in the following example. 

As usual, we define the geometry of the model by evaluating the following commands: 
\mathInTwoLines{<<xAct/xPand.m;}{DefManifold[ M, 4, \{$\alpha$, $\beta$, $\mu$, $\nu$\} ];}
\setcounter{inout}{2}
\mathInTwoLines{DefMetric[ -1, g[-$\mu$,  -$\nu$], CD, \{";", "CD"\} ];}{DefMetricPerturbation[ g, dg, $\epsilon$ ];}
\setcounter{inout}{4}
\mathInTwoLines{SetSlicing[ g, n, h, cd, \{"|", "D"\}, "FLFlat" ];}{order = 1;}
For simplicity, we customize the use of the command {\tt ToxPand} as 
\setcounter{inout}{6}
\mathIn{MyToxPand[\,expr$\_$,\,gauge$\_$,\,order$\_$\,]\,:=\,ToxPand[\,expr,\,dg,\,u,\,du,\,h,\,gauge,\,order\,];}
We define the constants $\kappa = 8 \pi G$ and $\Lambda$ with 
\mathInTwoLines{DefConstantSymbol[ $\kappa$ ];}{DefConstantSymbol[ $\Lambda$ ];}
and then the energy--momentum tensor of the fluid: 
\setcounter{inout}{9}
\mathIn{DefTensor[ T[-$\mu$, -$\nu$], M ]}
We introduce a boolean variable to switch on or off the pressure of the fluid filling the space--time. 
This allows us to consider either dust- or radiation-dominated era in a simple way. 
\mathIn{\$Dust\,=\,True;}
\mathIn{IndexSet[ T[$\alpha\_$,$\beta\_ $],\,%
	($\rho$u[]\,+\,If[\$Dust,\,0,\,Pu[]])\,u[$\alpha$]u[$\beta$]\,+\,If[\$Dust,\,0,\,Pu[]] g[$\alpha$,$\beta$]];}
The fields {\tt $\rho$u} (energy density), {\tt Pu} (pressure) and {\tt u} (fluid 4-velocity) will be automatically 
created later, when calling the function {\tt (My)ToxPand}. The user is encouraged to set the boolean variable to 
{\tt False} in order to include the effects of pressure. 

We now derive the perturbations of Einstein equations for the model at stake. We define the Einstein equations as 
\mathIn{MyGR[$\mu\_$, $\nu\_$] := EinsteinCD[$\mu$, $\nu$] + g[$\mu$,$\nu$] $\Lambda$/$\kappa$ - $\kappa$ T[$\mu$,$\nu$];}
and we evaluate their perturbations simply using 
\mathIn{MyGRresult = MyToxPand[ MyGR$[\mu,-\nu]$, "NewtonGauge", order ];}
Since the computation of this expression is the time-consuming part, we have stored its value in the variable {\tt MyGRresult}. 

The background components of the resulting expression are extracted using the \textit{xPand} function {\tt ExtractOrder[expr, 0]} 
(where $0$ indicates the background value), and the projected components with the \textit{xPand} function {\tt Extract\-Components}: 
\mathIn{ExtractComponents[\,ExtractOrder[ ah[]\^{}2 MyGRresult, 0 ], h, \{"Time","Time"\}\,]}
\mathOut{$\displaystyle \frac{\Lambda a^2}{\kappa} - 3{\cal H}^2 + \kappa \, a^2\rho $}
\mathIn{ExtractComponents[\,ExtractOrder[ah[]\^{}2 MyGRresult, 0], h, \{"Space","Space"\}\,]}
\mathOut{$\displaystyle \frac{\Lambda a^2 h^\mu_{\phantom{\mu} \nu}}{\kappa} 
	- h^\mu_{\phantom{\mu} \nu} \, {\cal H}^2 
	- 2 \, h^\mu_{\phantom{\mu} \nu} \, {\cal H}'$}
Similarly, the first-order equations are obtained by evaluating 
\mathIn{ExtractComponents[\,ExtractOrder[ ah[]\^{}2 MyGRresult, 1 ], h, \{"Time","Time"\}\,]}
\mathOut{$\kappa \, a^2\pertbis{1}{\rho}+ 6 \, {\cal H}^2\pertbis{1}{\phi} 
	+ 6 \, {\cal H} \pertbis{1}{\psi'} - 2D_\alpha D^\alpha \pertbis{1}{\phi}$}
\mathIn{ExtractComponents[\,ExtractOrder[ ah[]\^{}2 MyGRresult, 1 ], h, \{"Time","Space"\}\,]}
\mathOut{$-\kappa \, a^2\pertbis{1}{B_\nu} \, \rho - \kappa \, a^2\pertbis{1}{V_\nu} \, \rho 
	+ \frac{1}{2} \, D_\alpha D^\alpha \pertbis{1}{B_\nu} 
	- \kappa \,  a^2\rho \, D_{\nu}\pertbis{1}{V} 
	- 2 \, {\cal H}D_{\nu}\pertbis{1}{\phi} 
	- 2 \, D_{\nu}\pertbis{1}{\psi'}$}
\mathIn{ExtractComponents[\,ExtractOrder[ ah[]\^{}2 MyGRresult, 1 ], h, \{"Space","Time"\}\,]}
\mathOutTwoLines{$2\pertbis{1}{B^\mu}{\cal H}^2 
	- 2\pertbis{1}{B^\mu}{\cal H}' 
	+ \kappa \, a^2\pertbis{1}{V^\mu} \, \rho 
	+ D_\alpha \pertbis{1}{E'{}^{\mu\alpha}} 
	- \frac{1}{2} \, D_\alpha D^\alpha \pertbis{1}{B^\mu} 
	+ \kappa \, a^2\, \rho \, D^{\mu}\pertbis{1}{V}$}{$
	+ 2 \, {\cal H} \, D^{\mu} \pertbis{1}{\phi} + 2 \, D^{\mu}\pertbis{1}{\psi'}$}
\mathIn{ExtractComponents[\,ExtractOrder[ ah[]\^{}2 MyGRresult, 1 ], h, \{"Space","Space"\}\,]}
\mathOutThreeLines{$ \pertbis{1}{E^{''}{}^\mu_{\phantom{\mu} \nu}} 
	+ \pertbis{1}{E^{'}{}^\mu_{\phantom{\mu} \nu}} \, {\cal H} 
	+ 2 \, h^\mu_{\phantom{\mu} \nu} \, {\cal H}^2 \pertbis{1}{\phi} 
	+ 4 \, h^\mu_{\phantom{\mu} \nu} \, {\cal H}'\pertbis{1}{\phi'} 
	+ 2 \, h^\mu_{\phantom{\mu} \nu} \, {\cal H}\pertbis{1}{\phi'}
	+ 4 \, h^\mu_{\phantom{\mu} \nu} \, {\cal H}\pertbis{1}{\psi'} 
	+ 2 \, h^\mu_{\phantom{\mu} \nu} \pertbis{1}{\psi^{''}}$}{$%
	\;\; - D_{\alpha}D^{\alpha} \pertbis{1}{E^\mu_{\phantom{\mu} \nu}} 
	+ h^\mu_{\phantom{\mu} \nu} \, D_{\alpha}D^{\alpha}\pertbis{1}{\phi} 
	- h^\mu_{\phantom{\mu} \nu} \, D_{\alpha}D^{\alpha}\pertbis{1}{\psi} 
	- {\cal H} \, D^{\mu}\pertbis{1}{B_\nu} 
	- \frac{1}{2} \, D^{\mu}\pertbis{1}{B'_{\nu}} 
	- {\cal H} \, D_{\nu} \pertbis{1}{B^{\mu}}$}{$
	-\frac{1}{2} \, D_{\nu}\pertbis{1}{B'{}^{\mu}} 
	+ D_{\nu}D^{\mu}\pertbis{1}{\phi} 
	+ D_{\nu}D^{\mu}\pertbis{1}{\psi}$}
%

%%% Section
\section{Bianchi cosmologies} \label{sec:bianchi_cosmo}

%---Subsection
\subsection{Constructing a four-dimensional basis}

We here review the general properties of Bianchi space--times. The reader can find a detailed 
presentation of their classification in \cite{1969CMaPh..12..108E}, and summaries in, e.g., 
\cite{Ellis:1998ct,Pontzen:2007ii}. 

Bianchi space--times posses by definition a set of (three-dimensional) homogeneous hypersurfaces. 
One can therefore introduce three linearly independent spatial Killing vector fields (KVF) $\bm \xi_i$, 
with $i \in \{ 1, 2, 3 \}$, satisfying 
\be \label{eq:KVF}
	\Lie_{\bm \xi_i} \, \bar g_{\mu \nu} = 0 
		\;\, \Leftrightarrow \;\, 
			\bar \nabla_{(\mu} \, {\xi_i}_{\, \nu )} = 0 \, , 
		\qquad 
	\bar n_\mu \, \xi_i^\mu = 0 \, . 
\ee
From these properties together with equation \eqref{eq:vorticity}, we obtain 
\be \label{eq:KVF_prop}
	\Lie_{\bm{\bar n}}\, \bm \xi_i 
		= \left[ \bar{\gr{n}}, \bm \xi_i \right] 
		= 0 \, .  
\ee
The nature of the Bianchi model is determined by the spatial structure coefficients $C^k{}_{i j}$, 
defined from the commutators of the KVF: 
\be
	[\bm \xi_i, \bm \xi_j ] 
		\equiv - C^k_{\phantom{k} i j} \bm \xi_k \, , 
		\quad\; \mathrm{with} \quad 
	C^k_{\phantom{k} i j} = - C^k_{\phantom{k} j i} \, . 
	\label{eq:coeff_struct_xi}
\ee
The Jacobi identity 
\be
	\big[ \bm \xi_i, [ \bm \xi_j, \bm \xi_k ] \big] 
	+ \big[ \bm \xi_j, [ \bm \xi_k, \bm \xi_i ] \big] 
	+ \big[ \bm \xi_k, [ \bm \xi_i, \bm \xi_j ] \big] 
	= 0 \, , 
\ee
contrains these constants to verify
\be \label{eq:jacobi}
	C^m_{\phantom{m} [ i j} C^l_{\phantom{l} k ] m} = 0 
		\quad \Rightarrow \quad 
	C^m_{\phantom{m} i j} C^l_{\phantom{l} l m} = 0 \, , 
\ee
where we have used the fact that the $C^k{}_{i j}$ are constant on spatial slices. 

We now construct, on a given hypersurface, a vector basis $\{ {\bf e}_i \}$ and its dual $\{ {\bf e}^i \}$ 
invariant under the action of the KVF, namely satisfying: 
\be \label{eq:const_3basis}
	\Lie_{\bm \xi_i} {\bf e}_j = [\bm \xi_i, {\bf e}_j ] = 0 \, , 
		\qquad 
	\Lie_{\bm \xi_i} {\bf e}^j = 0 \, . 
\ee
From these properties along with relation \eqref{eq:coeff_struct_xi}, we infer that the $\bf{e}_i$ can be chosen such that 
\be
	[ {\bf e}_i, {\bf e}_j ] 
		= C^k_{\phantom{k} i j} {\bf e}_k \, , 
		\qquad 
	2 {e_i}^\mu {e_j}^\nu \nabla_{[\mu} {e^k}_{\nu]} 
		= - C^k_{\phantom{k} i j} \, . 
	\label{eq:coeff_struct_e}
\ee
The constants of structure can be further developed in terms of a symmetric `tensor' $N^{i j}$ 
and a `vector' $A^i$ as 
\be
	C^k_{\phantom{k} i j} 
		= \epsilon_{i j m} N^{m k} + 2 A_{[i} \delta^k_{\phantom{k} j]} 
	\label{eq:ToBianchiType}
\ee
where $\epsilon_{i j m}$ denotes the totally anti-symmetric Levi-Civita symbol. 
Note that the Jacobi identity~\eqref{eq:jacobi} translates in that case to the simple relation 
\be \label{eq:na}
	N_{ij} A^j = 0 \, . 
\ee
We then extend the bases $\{ {\bf e}_i \}$ and $\{{\bf e}^i \}$ to the whole space--time by Lie dragging 
them with $\bm{\bar n}$, which implies the properties 
\be \label{eq:const_4basis}
	\Lie_{\bm{\bar n}} {\bf e}_i = [\bm{\bar n}, {\bf e}_i ] = 0 \, , 
		\qquad 
	\Lie_{\bm{\bar n}} {\bf e}^i = 0 \, . 
\ee

With the above procedure, we are able to construct a four-dimensional basis 
$\{ {\bf e}_a \} \equiv \{ \bm{\bar n}, {\bf e}_i \}$ along with its dual 
$\{ {\bf e}^a \} \equiv \{ \bm{\underline{\bar n}}, {\bf e}^i \}$ 
(where $\bm{\underline{\bar n}}$ is the dual form of $\bm{\bar n}$ and $a \in \{ 0, 1, 2, 3 \}$),
that are invariant under the action of the KVF. 
The commutation relations of these new bases simply follow from expressions \eqref{eq:coeff_struct_e} and \eqref{eq:const_4basis}: 
the structure coefficients $C^c_{\phantom{c} a b}$ vanish when any of the indices is zero and take the values 
$C^k_{\phantom{k} i j}$ otherwise. 
This method to build a four-dimensional basis out of a three-dimensional one defined on a given spatial hypersurface 
is the simplest one\footnote{Note that in this framework, only $\bm{\bar n}$ is a unit vector. An alternative approach 
consists in building a basis of vectors that all are unitary, by renormalizing the ${\bf e}_i$. However, by doing so, 
the spatial vectors hence constructed do not commute with $\bm{\bar n}$ anymore, and their associated structure coefficients 
become time dependent. We shall not consider such possibility in the present paper, but details can be found in, e.g., 
\cite{1969CMaPh..12..108E,Ellis:1998ct,Pontzen:2007ii}.}. 

%---Subsection
\subsection{Expression of the geometrical tensors}

The components of the induced metric are written with respect to the Bianchi basis 
% $\{ {\bf e}_a \}$ 
$\{ {\bf e}_i \}$ 
and its dual as 
\be \label{eq:h_e}
	\bar h_{\mu\nu} 
		%= \bar h_{ab} \, {e^a}_\mu {e^b}_\nu
		= \bar h_{ij} \,{e^i}_\mu {e^j}_\nu \, . 
\ee
%where we have used the fact that ${\bf e}^0 = \bm{\underline{\bar n}}$ for the last equality. 
From relations \eqref{eq:KVF} and \eqref{eq:KVF_prop}, we deduce 
\be
	\Lie_{\bm \xi_i} \bar h_{\mu \nu} 
		= 0 \, ,
	\label{eq:lie_3met}
\ee
and using equation \eqref{eq:h_e} together with \eqref{eq:const_3basis}, we obtain: 
${\bf e}_k (\bar h_{i j}) = 0$. The components  $\bar h_{i j}$ are thus only time-dependent, 
namely $\bar h_{i j} = \bar h_{i j} (\eta)$. Any tensor field $\bm T$ lying on the 
background space--time possesses the same symmetries, and so we can expand it with respect to 
the Bianchi bases as 
\be
	\bar T_{\mu_1 \dots \mu_p} 
		= \bar T_{i_1\dots i_p}(\eta) \, {e^{i_1}}_{\mu_1} \dots  {e^{i_p}}_{\mu_p} \, .
\ee
Denoting by ${}^3\bar \Gamma_{ijk}$ the coefficients of the connection $\bar{\bm D}$ 
%in the bases $\{ {\bf e}_i \}$ and $\{{\bf e}^i \}$
in the Bianchi bases, we then deduce that 
\be \label{eq:cdOfBackgd}
	\bar D_k \bar T_{i_1\dots i_p} \equiv {e_k}^\alpha {e_{i_1}}^{\mu_1}\dots  {e_{i_p}}^{\mu_p} \bar D_\alpha \bar T_{\mu_1 \dots \mu_p} 
		= -\sum_{j=1}^p {{}^3 \bar \Gamma^{q}}_{k\, {i_j}} \bar T_{i_1 \dots i_{j-1} \,q  \,i_{j+1} \dots i_p}\,.
\ee
This relation is used in \textit{xPand} to compute the induced covariant derivative of 
any background tensor, such as the extrinsic curvature. 
The connection coefficients associated with the background connection $\bm{\bar \nabla}$ 
are given in the bases $\{ {\bf e}_a \}$ and $\{{\bf e}^a \}$ by
\be
	\bar \Gamma_{abc} 
		= \frac{1}{2} \big( - {\bf e}_a \left( \bar g_{bc} \right) + {\bf e}_b \left( \bar g_{c a} \right) + {\bf e}_c \left( \bar g_{ab} \right) 
		+ C_{a b c} - C_{b c a} + C_{c a b} \big) \, . 
\ee
Given that the components $\bar h_{ij}$ only depend on $\eta$, 
we deduce that the spatial-connection coefficients ${}^3 \bar \Gamma_{i j k}$ are expressed only in terms of the 
constants of structure. We have indeed 
\be \label{eq:GammaToC}
	{}^3 \bar \Gamma_{ijk} =\bar \Gamma_{ijk} 
		= \frac{1}{2} \big( C_{i j k} - C_{ j k i} + C_{k i j} \big) \, , 
		\qquad 
	C_{ijk} 
		= \bar \Gamma_{i j k}- \bar \Gamma_{i k j} \, .
\ee
Note that the tensor indices in the bases $\{ {\bf e}_i \}$ and $\{ {\bf e}^i \}$ are lowered with $\bar h_{ij}$ 
and raised with its inverse $\bar h^{ij}$, so that for instance $C_{kij}\equiv \bar h_{km} {C^m}_{ij}$. 
Note also that the constants of structure of the basis $\{ {\bf e}_i \}$, that we here note $C[{\bf e}]^k_{\phantom{k} ij}$ 
for the sake of clarity, are the components, in this specific basis, of a tensor. The associated components in a general 
basis can be recovered from\footnote{
Note that we are \textit{not} considering the constants of structure
of another basis with this construction. We rather build a tensor
${\bf C}[{\bf e}] \equiv C[{\bf e}]^k_{\,\,ij} {\bf e}_k \otimes {\bf
  e}^i \otimes {\bf e}^j$. It is obvious that the components of this
tensor in the basis  $\{ {\bf e}_i \}$ and $\{ {\bf e}^i \}$ are the
$C[{\bf e}]^k_{\,\,ij} $, but its components can also be taken in a
general basis even though we only refer to the commutation structure
of the ${\bf e}_i$. Since \textit{xTensor} manipulates only abstract indices
(that is indices in a general basis), this covariant point of view is necessary
to implement the structure of the Bianchi space--times in our package.
%{\color{red} XR: Maybe few words here to say that we are \textit{not} considering the constants of structure of another basis, as it is usually the case 
%(and hence our notation). Accordingly, the components $e$ may not need to be constant.}
%
} 
\be
	C[{\bf e}]^{\alpha}_{\phantom{\alpha} \mu\nu} 
		\equiv C[{\bf e}]^k_{\phantom{k} ij}\,{e_k}^\alpha {e^i}_\mu{ e^j}_\nu\,.
\ee
Similarly, the components of $N_{ij}$ and $A_i$ in a general basis are found from
\be
	N[{\bf e}]_{\mu\nu} 
		\equiv N[{\bf e}]_{ij} \, {e^i}_\mu{ e^j}_\nu \, , \quad 
	A[{\bf e}]_\mu \equiv A[{\bf e}]_i \, {e^i}_\mu\,.
\ee
Relation~\eqref{eq:cdOfBackgd} is also used to compute the induced covariant derivative 
of these three tensors, since they also live on the background space--time.  Note, finally, that 
from equation \eqref{eq:const_4basis} and the fact that the $C^i_{\phantom{i} jk}$ are constant, 
we obtain the useful relation 
\be \label{eq:LieC}
	\Lie_{\bm{\bar n}}({C^{\alpha}}_{\mu\nu}) = 0 \, .
\ee

The three-Riemann tensor of the hypersurfaces can be expressed only in terms of the constants of structure. 
In the bases $\{ {\bf e}_i \}$ and $\{ {\bf e}^i \}$, its components are given by
\begin{align} \label{eq:RiemannToConstants}
	{}^{3}\bar R_{ij}^{\phantom{ij} kl} =
		& - \frac{1}{2}{C^p}_{ij} {C_p}^{kl} + \frac{1}{2}C_{p\phantom{l}i}^{\phantom{p}l} {C^{pk}}_j 
		+ C_{p\phantom{l}j}^{\phantom{p}l} {C_i}^{kp}
		+ C_{p\phantom{l}j}^{\phantom{p}l} C^{k\phantom{j}p}_{\phantom{k}i} 
		+ {C}_{ijp} {C}^{pkl} 
		+ \frac{1}{2}C_{i\phantom{l}p}^{\phantom{i}l} {C_j}^{kp} 
		+ \frac{1}{2}{C^l}_{ip} C^{k\phantom{j}p}_{\phantom{k}j} 
		+ {C^k}_{jp} {C_i}^{lp} \, ,
\end{align}
%
%where we remind again that the indices on ${C^k}_{ij}$ are lowered and raised with
%$\bar h_{ij}$ and its inverse $\bar h^{ij}$, and 
where a double anti-symmetrization $[ij]$ and $[kl]$ is implied on the indices in the right-hand
side. The three-Ricci tensor and three-Ricci scalar can then be deduced, and we obtain
\begin{align}
	{}^3\bar R_{ij} = 
		& - \frac{1}{2} C_{kil} C^{k\phantom{j}l}_{\phantom{k}j}
		- \frac{1}{2} C_{kil} C^{l\phantom{j}k}_{\phantom{l}j} 
		+ \frac{1}{4} {C_i}^{kl} C_{jkl} 
		+ {C_{(ij)}}^p {C^k}_{pk} \, , \label{eq:RicciToConstants} \\
	{}^3\bar R = 
		& - \frac{1}{4} C_{ijk}C^{ijk} 
		- \frac{1}{2} C_{ijk} C^{jik} 
		+ {C^{kj}}_k {C^p}_{pj}\, . \label{eq:RicciScalToConstants}
\end{align}
Note that owing to the Jacobi identity~(\ref{eq:jacobi}), 
%the three-Riemann and three-Ricci tensors 
the three-curvature tensors 
can take several equivalent forms. Finally, due to the homogeneity of the hypersurfaces, 
any induced derivative acting on 
% ${}^{3}\bar R_{ijkl}$ 
the three-curvature tensors can be computed using equation~(\ref{eq:cdOfBackgd}). 
All the rules of this appendix, namely equations~(\ref{eq:cdOfBackgd}), (\ref{eq:GammaToC}), 
(\ref{eq:RiemannToConstants}), (\ref{eq:RicciToConstants}) and \eqref{eq:RicciScalToConstants}, 
are automatically created in \textit{xPand} when calling the function {\tt SetSlicing}, 
in case the specified space type is a Bianchi space--time. 
The boolean variable {\tt \$OpenConstantsOfStructure} controls whether or not the constants of structure should be opened 
in the final expressions using the parameterization~(\ref{eq:ToBianchiType}).

%%% Section
\section{Principal commands of xPand}

\vspace{-2mm}
\begin{table}[htb]
	\begin{tabular}{|c|c|c|}
		\hline
	{\tt Conformal}&
	{\tt ConformalWeight}&
	{\tt DefMatterFields} \\
	{\tt DefMetricFields}&
	{\tt \, DefProjectedTensor \,}&
	{\tt \, ExtractComponents \,} \\
	{\tt SetSlicing}&
	{\tt SplitGaugeChange}&
	{\tt SplitMatter} \\
	{\tt SplitMetric}&
	{\tt SplitPerturbations}&
	{\tt ToBianchiType} \\
	{\tt \, ToConstantsOfStructure \,}&
	{\tt ToxPand}&
	{\tt VisualizeTensor}\\[2pt]
                \hline
	\end{tabular}
	\caption{Information about these commands can be obtained within \textit{Mathematica} by evaluating {\tt ?NameOfCommand}.}
	\label{tab:commands}
\end{table}

\end{document}